\documentclass[10pt,aps,prb,twocolumn,nofootinbib,superscriptaddress,longbibliography]{revtex4-2}

\usepackage[utf8]{inputenc}
\usepackage{graphicx}
\usepackage{color,soul}
\usepackage{bm}
\usepackage{amsmath}
\usepackage{amssymb}
\usepackage{braket}
\usepackage[dvipsnames,svgnames,table]{xcolor}
\usepackage{fancyhdr}
\usepackage[english]{babel}
\usepackage{times}
\usepackage{soul}
\usepackage{hyperref}
\usepackage{mhchem}
\usepackage{multirow}

\usepackage{siunitx}

\hypersetup{
    unicode=true,		
    pdftoolbar=true,		
    pdfmenubar=true,		
    pdffitwindow=false,		
    pdfstartview={FitH},	
    pdfauthor={ELM},		
    colorlinks=true,		
    linkcolor=NavyBlue,		
    citecolor=Maroon,		
    filecolor=NavyBlue,		
    urlcolor=NavyBlue		
}

\newcommand{\tf}[1]{\text{#1}}

\begin{document}

\title{Position-dependent tight-binding model for Li impurities in monolayer and bilayer graphene}

\author{Hernán Aguirre}
\affiliation{Consejo Nacional de Investigaciones Cient\'{i}ficas y T\'{e}cnicas (CONICET), Instituto de Investigaciones en Fisicoqu\'{i}mica de C\'{o}rdoba (INFIQC), X5000HUA C\'{o}rdoba, Argentina}
\affiliation{Universidad Nacional de C\'{o}rdoba, Facultad de Ciencias Qu\'{i}micas, Departamento de Qu\'{i}mica Te\'{o}rica y Computacional, X5000HUA C\'{o}rdoba, Argentina}

\author{Hernán L. Calvo}
\affiliation{Instituto de F\'{i}sica Enrique Gaviola (IFEG-CONICET) and FaMAF, Universidad Nacional de C\'{o}rdoba, X5000HUA C\'{o}rdoba, Argentina}

\author{Eduardo M. Perassi}
\email{eduardo.perassi@unc.edu.ar}
\affiliation{Consejo Nacional de Investigaciones Cient\'{i}ficas y T\'{e}cnicas (CONICET), Instituto de Investigaciones en Fisicoqu\'{i}mica de C\'{o}rdoba (INFIQC), X5000HUA C\'{o}rdoba, Argentina}
\affiliation{Universidad Nacional de C\'{o}rdoba, Facultad de Ciencias Qu\'{i}micas, Departamento de Qu\'{i}mica Te\'{o}rica y Computacional, X5000HUA C\'{o}rdoba, Argentina}

\begin{abstract}
Lithium adsorption and intercalation can significantly modify the low-energy electronic properties of graphene-based materials, making their characterization relevant for understanding Li-ion transport and storage in graphitic electrodes. In this work, we investigate the electronic structure of a Li ion adsorbed on monolayer graphene (MLG) and intercalated within AB-stacked bilayer graphene (BLG). To this end, we develop a semi-empirical tight-binding model that incorporates Li-position dependence. Its parameters are determined by fitting to density-functional-theory calculations for different configurations, heights, and supercell sizes. The obtained model accurately reproduces the electronic bands near the Fermi level for both MLG and BLG and provides a transparent interpretation of the impurity-induced modifications in terms of symmetry breakings, intervalley mixing, and band-gap openings. We find that the perturbation introduced by the Li ion is strongly localized and that its effect decreases with increasing supercell size. The fitted parameters further reveal systematic differences between MLG and BLG in the spatial profile of the impurity potential. The obtained results provide an efficient framework for studying dilute Li impurities and constitute a useful starting point for future investigations of Li diffusion and impurity-induced transport phenomena in graphene-based materials.
\end{abstract}

\maketitle

\section{Introduction}

Lithium-ion batteries (LIBs) are the  dominant energy storage technology in diverse applications due to their high energy density, excellent coulombic efficiency, and low self-discharge rates. They are especially prevalent in automobiles and electronic devices  because they meet high energy demands.
In recent years, intensive research has focused on advancing LIB components---including anodes, cathodes, electrolytes, current collectors, and separators. All these efforts are  aimed at developing next-generation LIBs capable of meeting the requirements of applications such as electric vehicles, hybrid electric vehicles, aerospace applications, and autonomous electric devices that demand higher energy and power densities \cite{kim2019}. Li-ion battery technology is expected to remain central to energy storage for years to come \cite{durmus2020}.

To date, the state of the art in LIB electrodes is a  composite of materials that support ambipolar transport of Li ions and electrons. As the ionic and electronic conductivities of these electrodes increase, higher power densities can be achieved \cite{kuehne2017}.
Recently, graphene, a two-dimensional (2D) carbon allotrope composed of a single layer of carbon atoms arranged in a honeycomb lattice and the fundamental building block of other graphitic materials, has been considered  a promising electrode material for LIBs due to its high lithium storage capacity, high conductivity, and improved chemical and mechanical stability \cite{wang2009,thomas2018}.

Graphene exhibits unusual electronic properties such as massless Dirac fermions, the quantum Hall effect, an ambipolar electric-field effect, high carrier mobility at room temperature, and thermal conductivity as high as 4000 Wm$^{-1}$K$^{-1}$.
In addition, graphene has impressive mechanical and optical properties, including a fracture strength of 130 GPa, optical transparency, and an ultrabroad optical absorption spectrum \cite{sang2019}.
Bilayer graphene (BLG) also exhibits many properties similar to those of monolayer graphene (MLG), while introducing tunable electronic characteristics not present in the monolayer. These properties include excellent electrical  and thermal conductivities, with room-temperature mobility of up to $4\times10^4$ cm$^2$V$^{-1}$s$^{-1}$ and thermal conductivity of about 2800 Wm$^{-1}$K$^{-1}$, respectively; mechanical stiffness, strength, and flexibility, with a Young's modulus approaching 0.8 TPa; optical transparency with a white-light transmittance of approximately 95\%; and the possibility of chemical functionalization \cite{mccann2013}.

Recently, ultrafast lithium diffusion has been observed in BLG \cite{kuehne2017}. This discovery has stimulated growing interest in graphene as an electrode material for the next generation of LIBs. Materials with the exceptional electrical and ionic conductivity of graphene are essential for the efficient charge transport and energy storage required in LIB electrodes. Motivated by these experiments, in this work we investigate the interaction between electrons near the Fermi level and Li ions either adsorbed on MLG or intercalated within BLG. Specifically, we model how the electronic band structure near the Fermi level in both MLG and BLG is affected by the concentration and position of Li ions. The main goal of our proposed approach is to establish a generalized framework that captures the essential physics of electronic-ionic interactions, making it readily extensible to a wide range of applications, including electron-mediated ion transport through graphene-based nanodevices. This framework constitutes an important tool for studying phenomena such as ultrafast lithium diffusion \cite{kuehne2017,zhong2019}, as it allows the determination of current-induced forces (electromigration) \cite{todorov2014,bode2011} in the vicinity of impurities due to the electron flow across the device. To achieve this, we develop a Li-position-dependent tight-binding (TB) model based on the Slonczewski-Weiss-McClure (SWM) $\pi$-band framework, which is widely used in studies of graphite \cite{jung2013}. Unlike typical impurity models restricted to static, highly symmetric sites, our continuous pseudoorbital description keeps the number of fitting variables strictly constant and independent of the impurity position or supercell size. This approach aligns with continuous real-space parametrizations used to describe electrostatic perturbations from dopants \cite{lambin2012} and localized point defects \cite{fried2025}. By building upon these frameworks, our parameter-space efficiency prevents a growth in the number of parameters along continuous diffusion paths, offering a lightweight yet general framework for transport simulations. We particularize this model to evaluate how the Li-ion potential breaks symmetries in two benchmark cases: surface adsorption on MLG and interlayer intercalation within AB-stacked BLG.

The paper is organized as follows. Section~\ref{sec:method} outlines the first-principles methodology and the supercell configurations. Section~\ref{sec:TBmodel} presents the mathematical formulation of the position-dependent tight-binding model and introduces a partition scheme to analyze the energy bands at the supercell's $\Gamma$ point. Section~\ref{sec:results} discusses the electronic band structures and the impurity-induced effects for both MLG and BLG substrates. Finally, Section~\ref{sec:conclusions} summarizes the main conclusions and future outlooks of this work.

\section{Methodology}
\label{sec:method}
First-principles calculations based on density functional theory (DFT) were performed to obtain the electronic band structure of a Li ion adsorbed on MLG and intercalated within BLG. All calculations were carried out using the PWSCF code of the Quantum ESPRESSO distribution, employing the Perdew-Burke-Ernzerhof (PBE) exchange-correlation functional within the generalized gradient approximation (GGA) \cite{giannozzi2009,giannozzi2017}. Electron-ion interactions were described using PAW pseudopotentials for C atoms and an ultrasoft pseudopotential for Li, both generated within the PBE exchange-correlation functional. This combination provides an accurate description of the graphene electronic structure while allowing an efficient treatment of the Li valence electron. Valence-electron wave functions and charge density were expanded in a plane-wave basis set with kinetic-energy cutoffs of 45 Ry and 360 Ry, respectively.

\begin{figure}[t!]
  \centering
  \includegraphics[width=\linewidth]{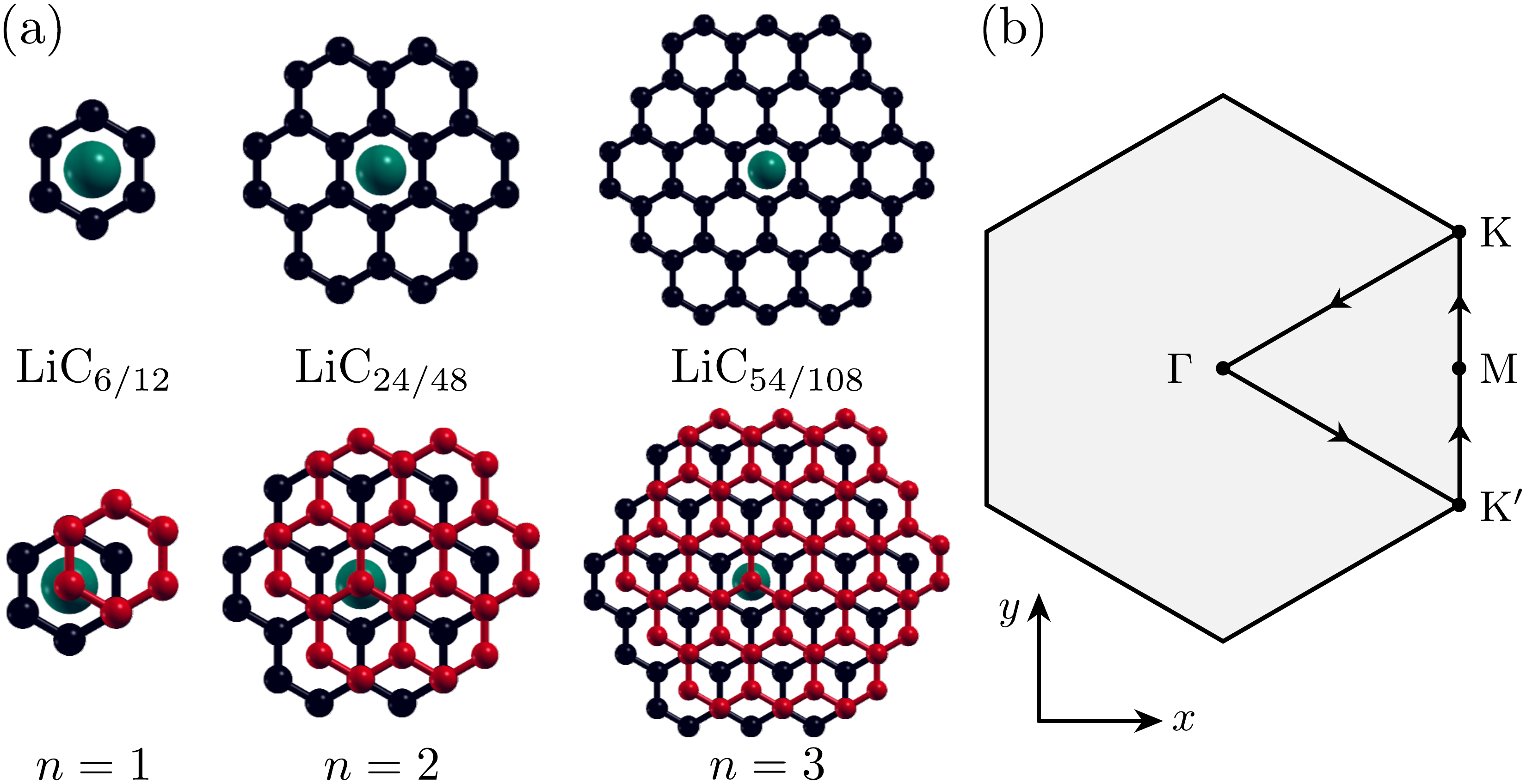}
  \caption{a) Three examples of MLG and BLG supercells of different sizes are shown, with $n = 1$ (6/12 C atoms), $n = 2$ (24/48 C atoms) and $n = 3$ (54/108 C atoms), and an adsorbed or intercalated Li ion, respectively. b) High-symmetry $\bm{k}$-points path (\ce{M-K-$\Gamma$-K$'$-M}) in the first Brillouin zone of the supercell.}
  \label{fig:supercells}    
\end{figure}

Kekul\'{e}-like supercells $(\sqrt{3}n\times\sqrt{3}n) \tf{R} 30^\circ $~\cite{andrade2025,farjam2009}, as shown in Fig.~\ref{fig:supercells}(a), were used for both MLG and BLG to model different Li dilution configurations.
The number of C atoms in the supercell is $6n^2$ for MLG and $12n^2$ for BLG, where $n \in \mathbb{N}$ and determines the length of the supercell side. Fig.~\ref{fig:supercells}(a) illustrates, from left to right, supercells corresponding to $n =$ 1, 2, and 3 for both MLG and BLG. Each supercell contains a single Li ion, generating a periodic structure for the composite system. Thus, the supercell size, given by $n$, determines the Li dilution configuration. The primitive lattice vectors of the supercell are given by:
\begin{equation}
  \bm{A}_1 =3na_0\frac{ \sqrt{3}\bm{e}_x+\bm{e}_y}{2}, \quad
  \bm{A}_2 =3na_0\frac{-\sqrt{3}\bm{e}_x+\bm{e}_y}{2},
\end{equation}
where $a_0 = 1.42$~{\AA} is the \ce{C-C} bond length.\footnote{For DFT calculations, $\bm{A}_3$ was chosen perpendicular to the graphene plane and sufficiently large to eliminate spurious interactions due to periodic boundary conditions.}
In the case of BLG, AB (Bernal) stacking was used with an interlayer spacing of 3.35~{\AA} \cite{zhang2021}.
The band structure was calculated along a high-symmetry path in the first Brillouin zone (FBZ) of the supercell, as depicted in Fig.~\ref{fig:supercells}(b).

To construct the TB model, a Hamiltonian was defined based on the atomic orbitals within the supercell shown in Fig.~\ref{fig:supercells}(a).
Only the $p_z$ orbitals of C atoms were included, while a single pseudoorbital was assigned to the Li ion.
The eigenvalues of the TB Hamiltonian were computed along the same $\bm{k}$-space path to obtain its band structure.
A custom code was developed for this purpose, utilizing LAPACK and the GNU Scientific Library (GSL) \cite{gough2009}.

To fit the TB parameters to a region of interest of the DFT band structure, a trust-region optimization method was employed.
Specifically, the Levenberg–Marquardt algorithm was utilized \cite{nocedal2006}. 
This algorithm was used to find the TB parameters, $\bm{\alpha}$, for each supercell size $n$.
A cost function, $\Phi_n(\bm{\alpha})$, was defined as follows:
\begin{equation}
\Phi_n(\bm{\alpha}) = \frac{1}{2}\sum_{i,j,s}\left[\omega_n(\bm{k}_j)f(\bm{\alpha},P_i,\bm{k}_j,s)\right]^2 \,, 
\label{eq:cost_function}
\end{equation}
where $f$ is the difference between the energy eigenvalues of the DFT and TB bands, $P_i$ labels the Li-ion position within the supercell, $\bm{k}_j$ are the $\bm{k}$-points along the $\bm{k}$-path, $s$ indicates the band, and $\omega_n$ is a statistical weight applied to $f$. This weight is defined as:
\begin{align}
\omega_n(g)= \begin{cases} 
1 & \text{if } |g| \leqslant \chi_0 \\
\beta & \text{if } \chi_0 < |g| \leqslant \chi_1 \\
0 & \text{if } |g|>\chi_1
\end{cases} \, ,
\label{eq:wn_f}
\end{align}
where $-1\leqslant g \leqslant 1$ linearly parametrizes the $\bm{k}$-path of Fig.~\ref{fig:supercells}(b), such that $\bm{k}=\bm{k}(g)$. Here, $\chi_0$, $\chi_1$, and $\beta$ are dimensionless parameters which may take different values depending on the size of the supercell. See Appendix~\ref{app:fitting} for further details regarding the fitting procedure.

\section{Position-Dependent Tight-Binding Model}
\label{sec:TBmodel}

\begin{figure}[t!]
  \centering
  \includegraphics[width=0.95\linewidth]{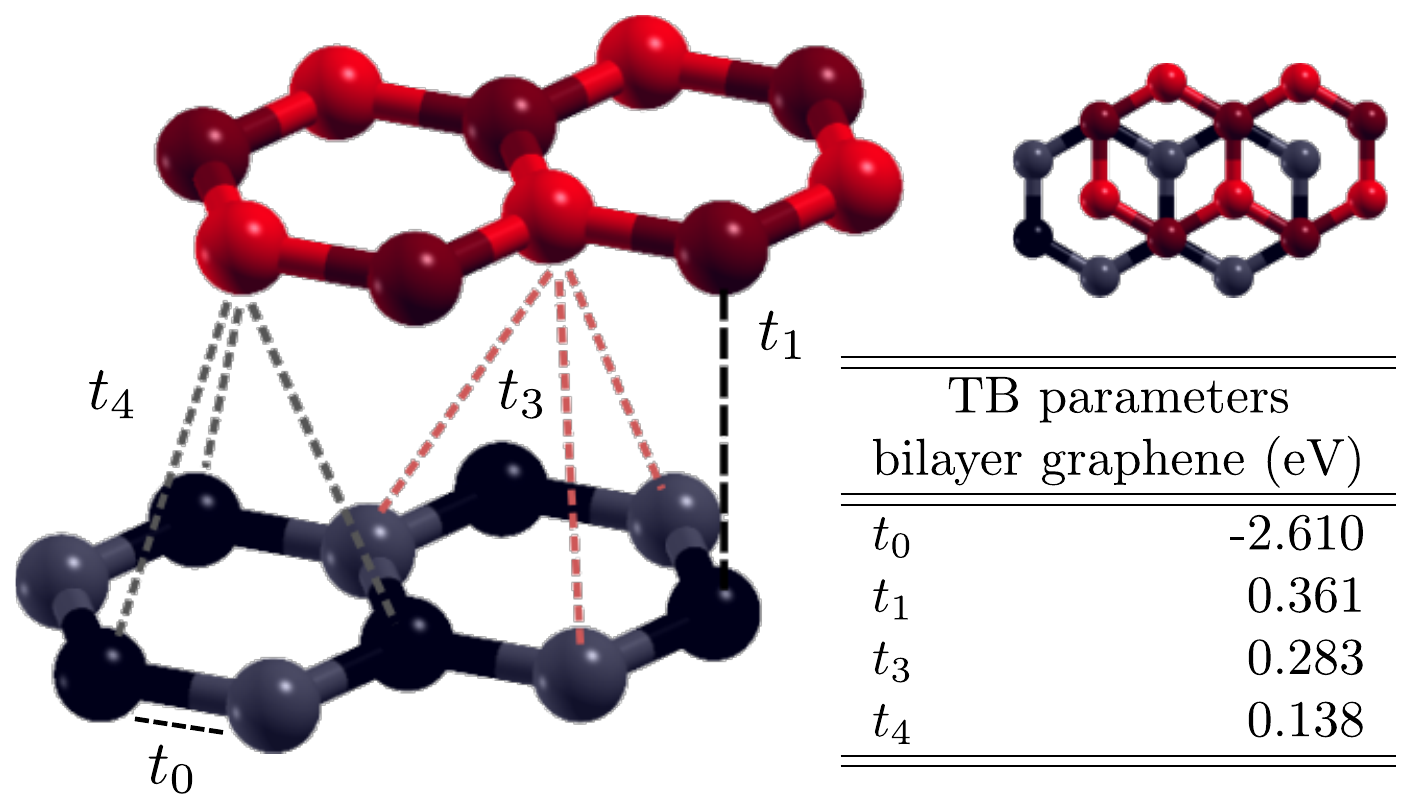}
  \caption{Schematics of the AB-stacked BLG and hopping amplitudes. Inset: Used tight-binding parameters for pristine BLG in eV units~\cite{jung2014}. For MLG, only $t_0$ is considered.}
  \label{fig:hoppings}
\end{figure}

It is well known that $\pi$-bands are responsible for the low-energy electronic properties of graphitic materials, including single and multilayer graphene \cite{jung2014}. Graphitic materials such as MLG and BLG have their Fermi levels at the Dirac points where two low-energy bands touch \cite{wallace1947,novoselov2005,castroneto2009,foatorres2014}. In order to study the interaction between electrons and Li ions around the Dirac points, it is convenient to use TB models based on orthogonal orbitals. To develop them, the Slonczewski-Weiss-McClure (SWM) model \cite{mcclure1957,slonczewski1958}, widely used for graphite, is adopted as a starting point in this work. Specifically, this work builds upon the formulation developed by Jung and MacDonald \cite{jung2014} for the pristine bilayer case. This model is characterized by a set of parameters that describe the band structure in a region close to the Dirac points, taking into account nearest-neighbor in-plane hoppings and interlayer couplings, as shown in Fig.~\ref{fig:hoppings}.
For MLG, the model is reduced to only considering the in-plane hopping $t_0$. Lithium is the third element of the periodic table and the lightest element that is solid at room temperature. It has two naturally occurring stable isotopes, \ce{^6Li} and \ce{^7Li}, with abundances of 7.4\% and 92.6\%, respectively \cite{tabelin2021}, and an electron configuration of $1s^22s^1$. Lithium is highly reactive and readily loses its single valence electron to form Li$^+$ cations.

To incorporate position dependence into the TB model, we introduce a pseudoorbital into the Jung and MacDonald framework to represent the Li ion on MLG and within BLG substrates. We assume an ellipsoidal shape to reflect the deformation of the lithium 2$s$ orbital, which is caused by its interaction with the substrate and results in anisotropy along the $z$-axis, perpendicular to the substrate plane. This pseudoorbital affects the onsite energies of its neighboring C atoms and is coupled to them by hopping terms. Both the onsite energies and the hoppings are functions of the \ce{Li-C} separation through a quadratic form which incorporates such pseudoorbital anisotropy.

The tight-binding Hamiltonian describing the composite system is written as $\hat{H} = \hat{H}_0 + \hat{H}_\tf{int}$, where $\hat{H}_0$ represents the Hamiltonian of the substrate (MLG or BLG) and $\hat{H}_\tf{int}$ accounts for the presence of the Li ions. In periodic substrates we have, in general:
\begin{equation}
\hat{H}_0 = \sum_{ij} \sum_{\alpha\beta} h^{(0)}_{\alpha\beta}(\bm{R}_i-\bm{R}_j) \hat{c}^\dagger_{\alpha,i} \hat{c}_{\beta,j},
\end{equation}
where the fermionic operator $\hat{c}^\dagger_{\alpha,i}$ ($\hat{c}_{\alpha,i}$) creates (annihilates) an electron at the $p_z$ orbital of the $\alpha$-th carbon atom within the supercell located at the Bravais lattice vector $\bm{R}_i$. In this description, $\alpha$ and $\beta$ label the carbon sites inside $i$ and $j$ supercells, respectively, while the matrix elements $h^{(0)}_{\alpha\beta}(\bm{R}_i-\bm{R}_j)$ account for the onsite energies and hopping amplitudes between these atoms.

The inclusion of a Li ion per supercell is accounted for by the following Hamiltonian:
\begin{widetext}
\begin{equation}
\hat{H}_\tf{int} = \epsilon_\tf{Li} \sum_i \hat{c}^\dagger_{\tf{Li},i} \hat{c}_{\tf{Li},i} + \sum_{ij} \sum_\alpha \epsilon(\bm{r}_{\alpha,i}-\bm{r}_{\tf{Li},j})  \hat{c}^\dagger_{\alpha,i} \hat{c}_{\alpha,i}
 + \sum_{ij} \sum_\alpha \gamma(\bm{r}_{\alpha,i}-\bm{r}_{\tf{Li},j}) ( \hat{c}^\dagger_{\tf{Li},j} \hat{c}_{\alpha,i} + \tf{h.c.} ),
 \label{eq:Hint}
\end{equation}
\end{widetext}
where $\hat{c}^\dagger_{\tf{Li},i}$ ($\hat{c}_{\tf{Li},i}$) creates (annihilates) an electron at the Li pseudoorbital of the $i$-th supercell, such that its position is given by $\bm{r}_{\tf{Li},i} = \bm{r}_\tf{Li} + \bm{R}_i$. The first term in this Hamiltonian describes the Li ions' onsite energy $\epsilon_\tf{Li}$, which will be taken as a model parameter to be adjusted. The second term gives a correction to the carbon onsite energies due to the presence of the Li ions. 
For each carbon orbital, these can be summed up over the Li ions in the lattice, yielding:
\begin{equation}
\Delta\epsilon(\bm{r}_{\alpha,i}) = \sum_j \epsilon(\bm{r}_{\alpha,i}-\bm{r}_{\tf{Li},j}) \, .
\label{eq:Li-pot}
\end{equation}  
Notice, in particular, that we define the positions of the carbon atoms as $\bm{r}_{\alpha,i} = \bm{r}_\alpha + \bm{R}_i$. Finally, the last term represents the \ce{Li-C} couplings. For the proposed model, we assume that both the onsite correction $\epsilon(\bm{r})$ and the hopping amplitude $\gamma(\bm{r})$ decay with the \ce{Li-C} separation according to the following Gaussian form:
\begin{equation}
\epsilon(\bm{r}) = \epsilon_0 e^{-\sum_i (x_i/\sigma_i)^2}, \quad 
\gamma(\bm{r})   = \gamma_0   e^{-\sum_i (x_i/\tau_i)^2},
\label{eq:explicit}
\end{equation}
where $x_i = \bm{r}\cdot\bm{e}_i$ denote the cartesian components of $\bm{r}$, while $\bm{\sigma}$ and $\bm{\tau}$ introduce the associated anisotropies of the model. Such continuous Gaussian profiles are well established for modeling graphene defects, capturing both long-range electrostatic perturbations \cite{lambin2012} and localized interactions \cite{fried2025}. For our purposes, we will assume axial symmetry of the Li pseudoorbital along the $z$ direction, such that $\sigma_x = \sigma_y = \sigma$ and $\tau_x = \tau_y = \tau$. This yields 7 free parameters to be adjusted: 
\begin{equation}
  \bm{\alpha} = (\epsilon_\tf{Li},\epsilon_0,\gamma_0,\sigma,\sigma_z,\tau,\tau_z).
  \label{eq:parameters}
\end{equation}
To compare both DFT and TB energy bands, we compute the Bloch Hamiltonian associated with the proposed TB model. Noting that  
$\hat{H}$ is translationally invariant with respect to the lattice vectors, we can write it in the following general form:
\begin{equation}
\hat{H} = \sum_{ij}\sum_{\alpha\beta} h_{\alpha\beta}(\bm{R}_i-\bm{R}_j) \hat{c}^\dagger_{\alpha,i} \hat{c}_{\beta,j} \, ,
\end{equation}
where the $\alpha$ and $\beta$ indices also include the Li-ion pseudoorbital. By expressing the above operators in their Fourier representation:
\begin{equation}
\hat{c}^\dagger_{\alpha,i} = \frac{1}{\sqrt{N}} \sum_{\bm{k}} e^{i\bm{k}\cdot\bm{R}_i} \hat{c}^\dagger_\alpha(\bm{k}) \, ,
\end{equation} 
with $N$ the number of supercells, and using the periodic condition on the matrix elements $h_{\alpha\beta}$, we arrive at the following expression:
\begin{align}
\hat{H} &= \sum_{\bm{k}} \sum_{\alpha\beta} h_{\alpha\beta}(\bm{k}) \hat{c}^\dagger_\alpha(\bm{k}) \hat{c}_\beta(\bm{k}), \\ 
h_{\alpha\beta}(\bm{k}) &= \sum_{\bm{R}} h_{\alpha\beta}(\bm{R}) e^{i\bm{k}\cdot\bm{R}} \, .
\end{align}
The construction of the Bloch Hamiltonian through $h_{\alpha\beta}(\bm{k})$ thus reduces to the computation of the matrix elements $h_{\alpha\beta}(\bm{R})$. Since the \ce{Li-C} hopping terms are assumed to rapidly decay with the in-plane distance, it is enough to consider nearest-neighbor supercells, meaning that the possible values for the lattice vector $\bm{R}$ are $\{\bm{0}, \pm \bm{A}_1, \pm \bm{A}_2, \pm (\bm{A}_1 + \bm{A}_2 )\}$.
Importantly, the model is general for impurities arranged regularly on a periodic lattice, forming a composite system that establishes a well-defined crystal structure. For the present work, this means we can simply switch $\hat{H}_0$ between MLG and BLG, taking the proper Hilbert space and C positions. In either case, one should obtain different values for the same set of fitting parameters.\\

\textit{Partition scheme for the effective Hamiltonian}. To further analyze the modifications introduced by the Li ion on the energy bands around the Dirac points, we construct an effective Hamiltonian at the supercell $\Gamma$ point, following the partitioning technique discussed in Appendix~\ref{app:effective_hamiltonian}. We denote the pristine MLG or BLG eigenstates as $\ket{\bm{k},s}$, where $s$ labels the valence and conduction bands. We focus on the fourfold degenerate subspace $\mathcal{D}$ at the supercell $\Gamma$ point, which arises from the folding of the $\bm{K}$ and $\bm{K}'$ valleys of the pristine substrate.\footnote{For BLG, the low-energy subspace near the Fermi level is also fourfold degenerate, as we focus on the two bands closest to the Dirac point per valley.} By defining $\mathcal{C}$ as the complementary subspace spanned by the lithium orbital and all remaining substrate states folded into the $\Gamma$ point, the total Hamiltonian can be projected onto $\mathcal{D}$ as:
\begin{equation}
  \bm{h}_{\mathcal{DD}}^\tf{eff}(\epsilon) = \bm{h}_{\mathcal{DD}}+\bm{\Sigma}_{\mathcal{DD}}(\epsilon),
  \label{eq:Heff}
\end{equation}
with
\begin{equation}
  \bm{\Sigma}_{\mathcal{DD}}(\epsilon) =\bm{h}_{\mathcal{DC}} (\epsilon-\bm{h}_{\mathcal{CC}})^{-1}  \bm{h}_{\mathcal{CD}},
\end{equation}
where $\bm{h}_{\mathcal{DC}}$ and $\bm{h}_{\mathcal{CD}}$ are the coupling matrices between both subspaces. The resulting eigenvalue equation, $\bm{h}_{\mathcal{DD}}^\tf{eff}(\epsilon)\bm{\psi}_{\mathcal{D}} = \epsilon \bm{\psi}_{\mathcal{D}}$, is inherently nonlinear due to the energy dependence of the self-energy term $\bm{\Sigma}_{\mathcal{DD}}(\epsilon)$. To simplify the analysis, we evaluate the self-energy at a constant energy reference, approximating $\epsilon$ by the average potential energy introduced by the Li ion [Eq.~\eqref{eq:Li-pot}], i.e., $\epsilon = \bar{\Delta}$. Within this basis, the effective Hamiltonian allows us to distinguish two distinct types of corrections near the Dirac points:
\begin{itemize}
\item A \textit{direct} correction, $\bm{h}_\mathcal{DD}$, whose diagonal elements represent the average electrostatic potential exerted by the Li ion on the C atoms, while its off-diagonal elements introduce intravalley and intervalley couplings.
\item An \textit{indirect} correction, $\bm{\Sigma}_\mathcal{DD}$, which describes the effective coupling between the states in $\mathcal{D}$ mediated by the virtual transitions through the states in $\mathcal{C}$, including the Li atomic orbital.
\end{itemize}
Regarding the direct term, we identify the following specific matrix structure:
\begin{equation}
\bm{h}_\mathcal{DD} = \bar{\Delta} \times \bm{1}_\mathcal{D}+
\begin{pmatrix}
0 & v & w & w_\tf{x} \\
v & 0 & w_\tf{x} & w \\
w^* & w_\tf{x}^* & 0 & v \\
w_\tf{x}^* & w^* & v & 0
\end{pmatrix},
\label{eq:hdd}
\end{equation}
where $v$ represents the real-valued interband-intravalley coupling, while $w$ and $w_\tf{x}$ correspond to the complex intraband-intervalley and interband-intervalley couplings, respectively, with the asterisk denoting complex conjugation. Explicit expressions for these coupling matrix elements for both MLG and BLG are provided in Appendix~\ref{app:effective_hamiltonian}. The analytical eigenvalues of $\bm{h}_\mathcal{DD}$ are given by:
\begin{equation}
\lambda_\mathcal{D} = \left\{ \bar{\Delta} + v \pm |w + w_\tf{x}|, \, \bar{\Delta} - v \pm |w - w_\tf{x}| \right\}.
\end{equation}
Consequently, the lifting of the valley and band degeneracies is determined by the spatial dependence of these coupling terms as the Li ion moves across the supercell.

In general, the indirect correction matrix $\bm{\Sigma}_\mathcal{DD}$ cannot be reduced to the simple form of Eq.~\eqref{eq:hdd}. However, because it scales inversely with the energy separation between the $\mathcal{D}$ and $\mathcal{C}$ states, its matrix elements are typically smaller than those of $\bm{h}_\mathcal{DD}$. Despite their smaller magnitude, these indirect contributions can induce key band structure modifications when the direct term reduces to a purely diagonal matrix, i.e., $\bm{h}_\mathcal{DD} = \bar{\Delta} \bm{1}_\mathcal{D}$. Importantly, while $\bm{h}_\mathcal{DD}^\tf{eff}$ successfully links the local potential of the Li ion to the \textit{qualitative} behavior of the energy levels at the Dirac points, an accurate quantitative assessment of the final band energies still requires the full diagonalization of the Bloch Hamiltonian.

\section{Results and discussion}
\label{sec:results}

\begin{figure}[t!]
  \centering
  \includegraphics[width=\linewidth]{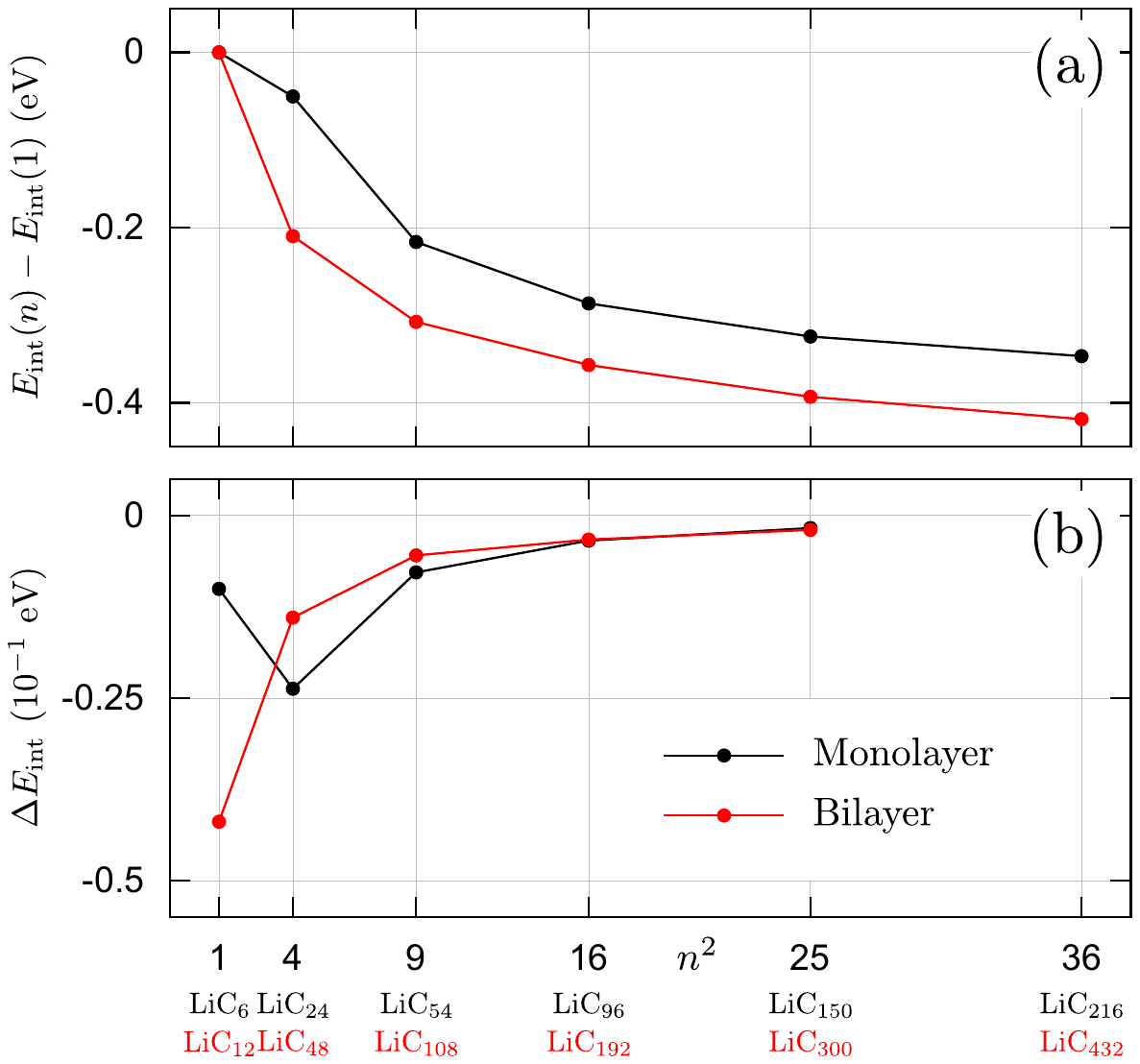}
  \caption{Interaction energy $E_\tf{int}$ (a) and corresponding slope $\Delta E_\tf{int}$ (b) as a function of the supercell size for MLG (black) and BLG (red) substrates, respectively.}
  \label{fig:bond_energy}
\end{figure}

Before applying the model to describe the electronic band structure around the Fermi level of the systems, the energy contributions of the \ce{Li-C} and \ce{Li-Li} interactions to the composite system were studied. In Fig.~\ref{fig:bond_energy} we show the energy associated with a Li ion on MLG and BLG as a function of the supercell size ($n^2$), yielding different Li concentrations, as illustrated in Fig.~\ref{fig:supercells}. The energy values are computed by using DFT at the most stable adsorption site for Li, and defined by $E_\tf{int} = E_{\tf{Li+s}}-E_\tf{s}$; where the interaction energy $E_\tf{int}$ accounts for both \ce{Li-Li} and \ce{Li-C} interactions, $E_\tf{Li+s}$ is the total energy of the Li-substrate composite system, and $E_\tf{s}$ denotes the energy of the substrate in the absence of the Li ion.
In Fig.~\ref{fig:bond_energy}(a), the reference energy (zero point) is taken as the energy corresponding to $n=1$. A comparison reveals that \ce{Li-Li} interactions are stronger on MLG than within BLG. This difference becomes increasingly evident as $n$ increases from 1 to 6, where the energy in BLG decreases more rapidly with increasing supercell size, as compared to MLG. This behavior may be attributed to the second layer above the Li ion, which likely screens the \ce{Li-Li} interaction \cite{wang2007,pollak2010,astles2024}.

Additionally, these curves asymptotically converge for larger supercell sizes, which evidences the decay of the \ce{Li-Li} interaction. At this limit, the system is dominated by the interaction between the Li ion and the substrate. To further highlight this behavior, we show in Fig.~\ref{fig:bond_energy}(b) the slope of these energy curves, i.e.,
\begin{equation}
\Delta E_\tf{int}(n) = \frac{E_\tf{int}(n+1)-E_\tf{int}(n)}{2n+1}.
\end{equation}
By comparing them, a distinct behavior is observed for small supercell sizes: for $n=1$, the energy decay slope in BLG is significantly steeper, though it converges faster than that of MLG. This difference already disappears from $n=4$ onward, at which point the two curves superimpose.
The slower convergence on MLG may be attributed to the absence of \ce{Li-Li} interaction shielding, which promotes the formation of a weakly adsorbed two-dimensional Li lattice on the substrate \cite{pollak2010}.
Based on this study, the model was applied to describe the different states of Li dilution on MLG and BLG, where both the \ce{Li-Li} and \ce{Li-C} interactions depend on the specific supercell size $n$. Although Fig.~\ref{fig:bond_energy} shows the calculated energies for supercell sizes up to $n=6$, band-structure calculations of the next section require significantly larger computational resources and were therefore limited to sizes up to $n = 4$ for MLG and $n = 3$ for BLG.

\subsection{DFT and TB energy bands}

\begin{figure}[t!]
  \centering
  \includegraphics[width=\linewidth]{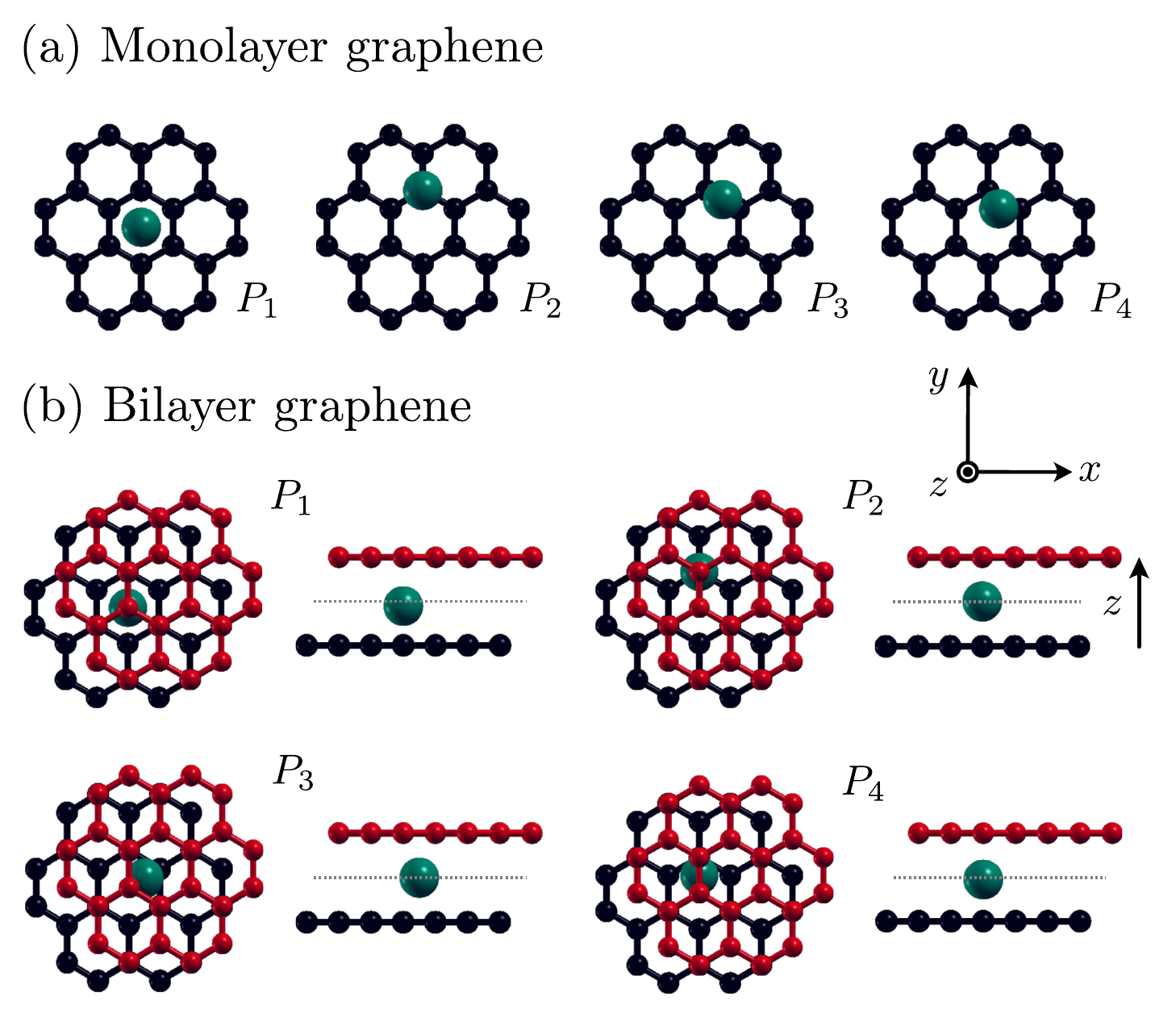}
  \caption{Schematic representation of the considered positions $P_i$ of the Li-ion adsorbed on MLG (a) and intercalated within BLG (b). The dotted gray lines in panel b indicate the midplane between the two graphene sheets.}
  \label{fig:Li_positions}
\end{figure}

In Fig.~\ref{fig:Li_positions} we display the different configurations for a Li ion adsorbed on MLG (a) and intercalated within BLG (b) substrates, as illustrated for a supercell with $n = 2$. These configurations were also considered for $n=3$ and $4$ in MLG and for $n=3$ in BLG.

In MLG, the lowest-energy configuration corresponds to the position labeled $P_1$ (hollow), with six-fold rotational symmetry, $C_6$, and three symmetry planes. At this position, the Li ion is located at a distance of 1.6857~{\AA} above the substrate plane. The other positions, labeled $P_2$, $P_3$, and $P_4$, were selected in decreasing order of local symmetry. For example, $P_2$ (top) and $P_3$ (bond) exhibit $C_3$ and $C_2$ rotational symmetries and three and two symmetry planes, respectively. The final position, $P_4$, retains only one symmetry plane.

For all considered positions, the Li ion is placed at a fixed vertical distance of 1.6857~{\AA} above the carbon layer, corresponding to the lowest-energy configuration (position $P_1$). Relative to $P_1$, the remaining configurations increase in energy in the order of $P_4$, $P_3$, and $P_2$.
We also performed energy-band calculations for the same in-plane positions, considering two additional Li heights of 1.8857~{\AA} and 1.4857~{\AA}. These configurations allow us to explore the electronic response of the system across a three-dimensional region surrounding the energy minimum.

For BLG [see Fig.~\ref{fig:Li_positions}(b)] we also considered different positions of the Li ion. For example, the lowest-energy configuration is $P_1$ (hollow-top), where the Li ion is located 1.4966~{\AA} above the center of the bottom layer (hollow) and below a C atom of the top layer (top).\footnote{An interlayer separation of 3.35~{\AA} was adopted for the DFT and TB calculations.} This configuration exhibits $C_3$ rotational symmetry and three symmetry planes.
In the highest-energy configuration, $P_2$ (top-top), the Li ion is placed midway between the two layers, directly above a carbon atom of the bottom layer and below another carbon atom of the top layer, yielding the same symmetry as $P_1$.
The remaining configurations, $P_3$ and $P_4$, take intermediate energy values and show $C_2$ rotational symmetry and a symmetry plane, respectively. For $P_3$, the Li ion is midway between the two layers, whereas for $P_4$ it is located 1.5858~{\AA} above the bottom layer.\\

\begin{figure}[t!]
  \centering
  \includegraphics[width=\linewidth]{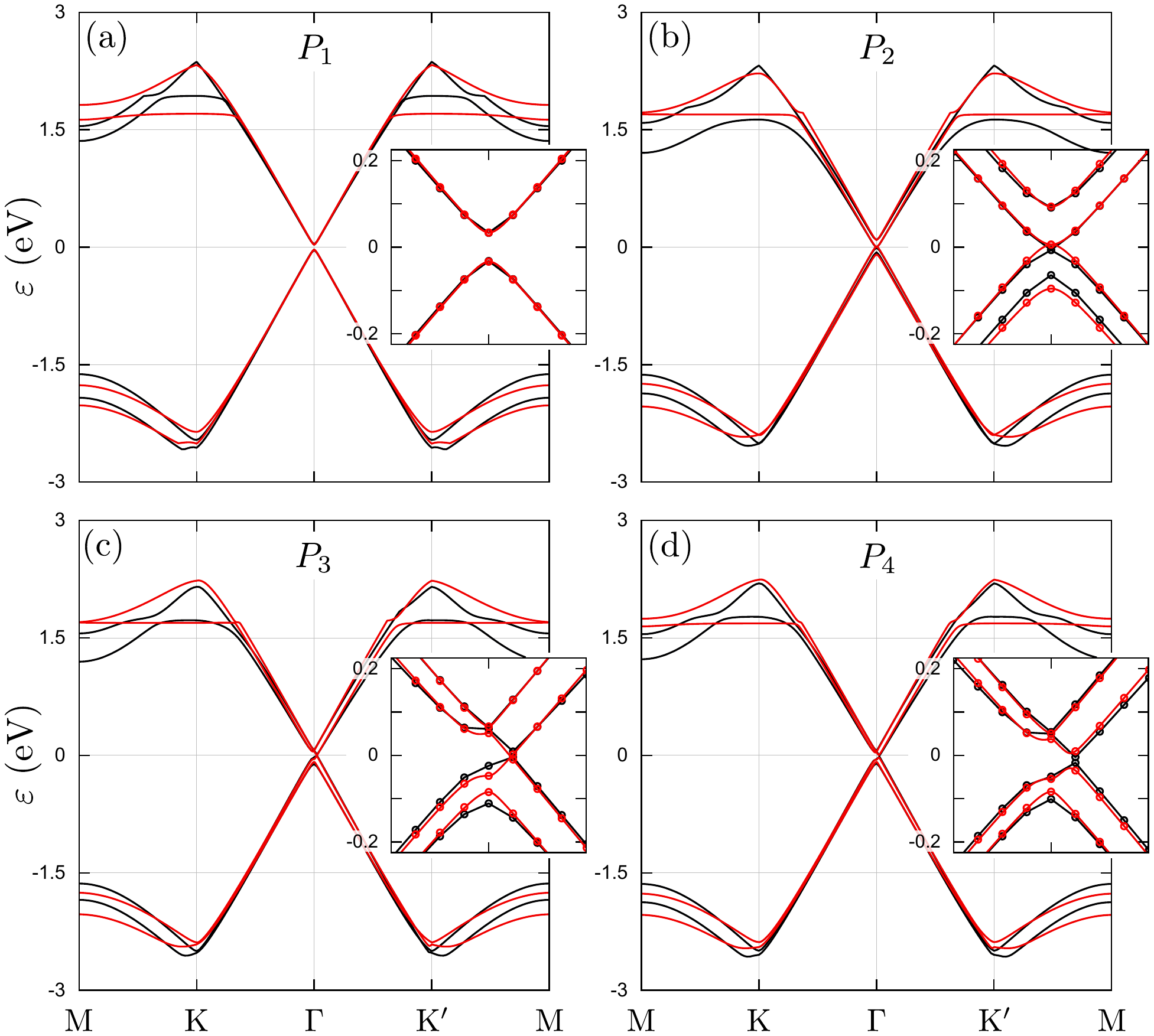}
  \caption{Energy spectra for Li adsorbed on MLG, relative to the Fermi level (Dirac point) of the pristine substrate. Here we take $n=2$ and the Li-ion positions depicted in Fig.~\ref{fig:Li_positions}(a).}
  \label{fig:bands_mono_pos}
\end{figure}

\textit{Monolayer graphene}. In Fig.~\ref{fig:bands_mono_pos} we show the DFT-calculated energy bands near the Fermi level, represented by black lines, for Li on MLG with a supercell size of $n = 2$. Each panel corresponds to a particular Li position within the supercell, see Fig.~\ref{fig:Li_positions}(a). The figure also shows the adjusted energy bands (red lines) from the TB model described in Sec.~\ref{sec:TBmodel}.

In Fig.~\ref{fig:bands_mono_pos}(a) we show the four electronic bands near the Fermi level for the lowest-energy configuration labeled $P_1$ in Fig.~\ref{fig:Li_positions}(a). The $\Gamma$ point corresponds to the  $\tf{K}$ and $\tf{K}'$ points of the pristine MLG folded into the supercell FBZ. From the figure, it can be seen that the TB model reproduces most of the DFT band features. The agreement is particularly strong near the $\Gamma$ point, as this region was prioritized during the fitting process via the weight function $\omega_n$ of Eq.~\eqref{eq:wn_f}, see Appendix~\ref{app:fitting}. For the chosen path, the bands exhibit mirror symmetry around the $\Gamma$ point. While near this point the bands remain almost two-fold degenerate, the presence of the Li ion removes the valley degeneracy for general $k$-directions and also opens a band gap of $\Delta = 65$ meV around the Fermi energy, which is shown in the inset of the figure. This is mainly attributed to the local potential introduced by the Li ion on the C atoms, which breaks the equivalence between the six C atoms in the surrounding hexagon and the remaining 18 at the supercell boundary.

In the calculation of the effective Hamiltonian $\bm{h}_\mathcal{DD}^\tf{eff}$ of Eq.~\eqref{eq:Heff} for position $P_1$, we find no direct band or valley mixing terms in the block $\bm{h}_\mathcal{DD}$, i.e., $v=w=w_\tf{x}=0$. Therefore, the band gap opening is accounted for by an effective intervalley coupling due to the indirect correction $\bm{\Sigma}_\mathcal{DD}$, which breaks the fourfold degeneracy into two doubly degenerate eigenenergies.

Fig.~\ref{fig:bands_mono_pos}(b) shows the energy bands for position $P_2$, where the Li ion is located directly above one C atom. The four bands are well separated around the $\Gamma$ point; however, no gap opens at the Fermi level in this specific case. The band separation is already obtained from the direct correction $\bm{h}_\mathcal{DD}$, and can be related to the broken sublattice symmetry. Although not shown, the lowest and highest energy bands exhibit a higher weight on the sublattice closer to the Li ion. Conversely, the intermediate bands show a higher weight on the other sublattice. 

Regarding the effective Hamiltonian in Eq.~\eqref{eq:Heff}, we find that the intervalley couplings satisfy $w_\tf{x}=-w$. Hence, we obtain four eigenvalues that split into three distinct energy levels: $\lambda_\mathcal{D} = \{\bar{\Delta}+v, \, \bar{\Delta}-v \pm 2|w|\}$. The first level is doubly degenerate and, since we obtain $|v|<|w|$, it is located between the other two. Furthermore, the indirect term $\bm{\Sigma}_\mathcal{DD}$ acquires the same structure as the matrix $\bm{h}_\mathcal{DD}$ in Eq.~\eqref{eq:hdd}, preserving the relation $w_\tf{x}=-w$ for the indirect intervalley couplings.

Comparing the TB model with the DFT calculation, we observe that the former yields a better agreement for the separation between the two upper bands than for that between the two lower bands. This can be related to the fact that the TB model only takes into account $p_{z}$ orbitals for the C atoms, thus yielding a more symmetric structure around the Dirac points, compared to the DFT calculation, which includes all C valence orbitals.

For position $P_3$, where the Li ion is located above a \ce{C-C} bond [Fig.~\ref{fig:bands_mono_pos}(c)], the four bands are separated, but two of them cross in the vicinity of the $\Gamma$ point. The bands are no longer symmetric along this path, as both the $x$- and $y$-mirror symmetries are broken by the chosen bond. In this case, the TB model faithfully reproduces the features near the $\Gamma$ point, with higher accuracy in the conduction bands.

Interestingly, for this particular configuration, the two Dirac points reappear;\footnote{For the chosen path, only one of the Dirac points is visible in the figure.} however, they are displaced from their original positions along the bond direction. Crucially, they remain connected through time-reversal symmetry; i.e., if $\bm{k}_0$ denotes the position of one Dirac point, the other is located at $-\bm{k}_0$. Regarding Eq.~\eqref{eq:hdd}, we find that $v=0$, as this configuration preserves sublattice symmetry. However, the intervalley couplings $w$ and $w_\tf{x}$ are different, meaning that the direct correction alone breaks the fourfold degeneracy into four distinct eigenenergies. The matrix elements of the indirect correction, on the other hand, are one order of magnitude smaller than those of the direct term, so only small deviations from the direct eigenvalues are expected.

Finally, Fig.~\ref{fig:bands_mono_pos}(d) corresponds to position $P_4$, where the Li ion is located at the center of one of the triangles forming the graphene hexagon. Here, due to its proximity to position $P_3$, the energy bands develop a structure similar to that of panel c, although in this case the Li-ion potential breaks the necessary symmetry and introduces a band gap. The TB model again shows good agreement with the calculated DFT bands on average, particularly near the $\Gamma$ point. Interestingly, for positions $P_3$ and $P_4$, the direct correction splits the fourfold degenerate subspace into two doubly degenerate bands, whereas the indirect correction lifts this remaining degeneracy. 

Importantly, since the system preserves time-reversal symmetry, the energy bands obey $\epsilon(\bm{k}) = \epsilon(-\bm{k})$, resulting in an identical band profile along the $\ce{M-K}$ and $\ce{M-K}'$ path segments for all Li-ion positions studied. Nevertheless, local electronic features near the $\Gamma$ point reflect the specific symmetry breaking induced by the Li-ion position.\\

\begin{figure}[t!]
  \centering
  \includegraphics[width=\linewidth]{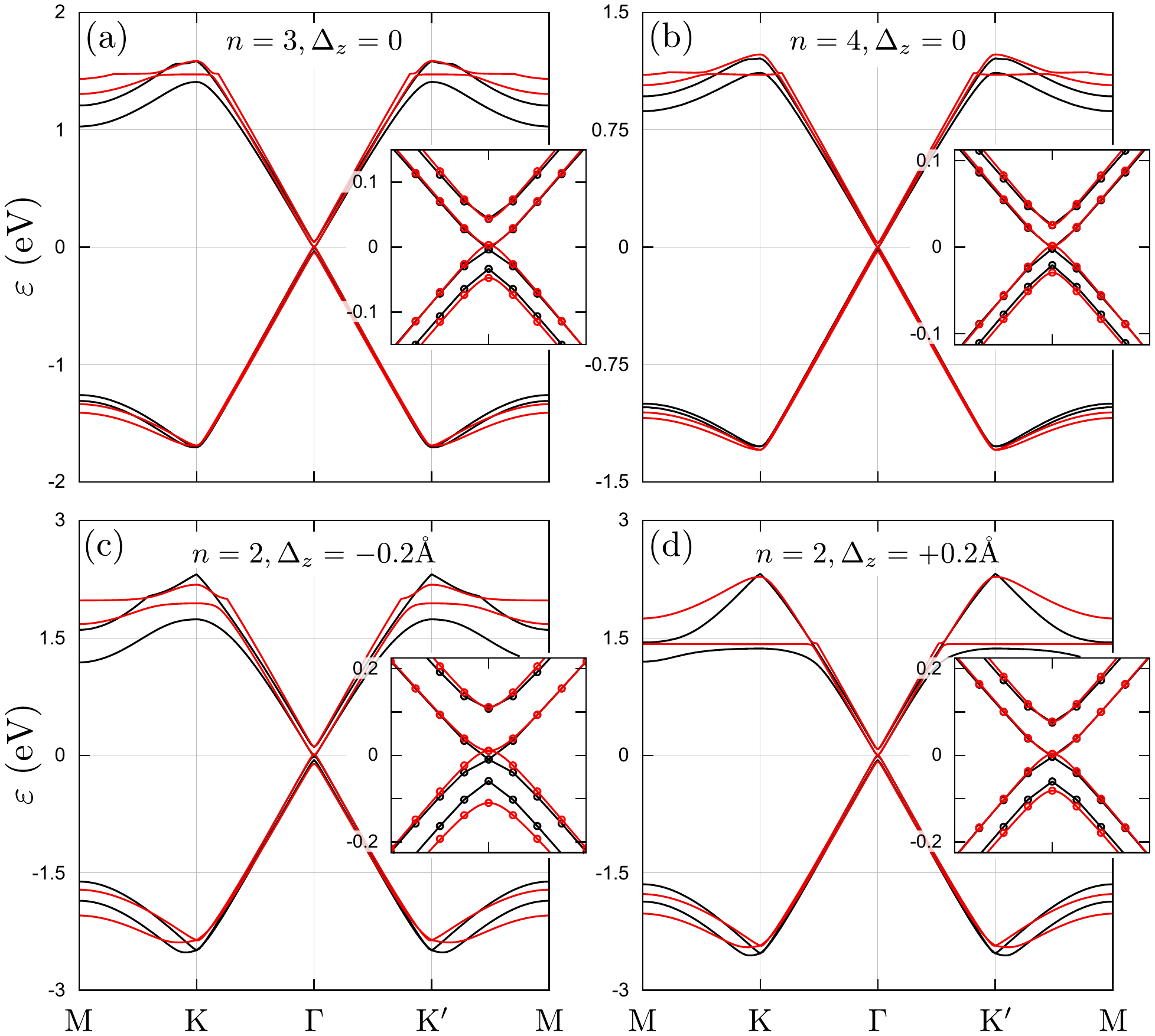}
  \caption{Band structure dependence on the supercell size and Li height on MLG at position $P_2$. Here we use $n=3$ (a) and $n=4$ (b) for $z=1.6857$~{\AA} ($\Delta_z = 0$), while in panels (c) and (d) we use $z=1.4857$~{\AA} ($\Delta_z = -0.2$~{\AA}) and $z=1.8857$~{\AA} ($\Delta_z = +0.2$~{\AA}) for $n=2$, respectively.  DFT bands are depicted in black, and the adjusted TB bands in red, respectively.}
  \label{fig:bands_mono_size}
\end{figure}

\textit{Supercell size and Li height effects}. Figs.~\ref{fig:bands_mono_size}(a) and (b) illustrate the band structure behavior for the Li-ion position $P_2$ across different supercell sizes: $n=3$ (a) and $n=4$ (b). In these panels, the electronic bands exhibit the same qualitative features previously discussed for $n=2$ and shown in Fig.~\ref{fig:bands_mono_pos}(b).
However, the energy scale at which these features are visible is reduced as $n$ increases. This is the expected trend, as the potential is short-ranged; thus, with larger $n$, an increasing number of C atoms in the supercell remain unaffected (or only weakly perturbed) by the Li ion. In this sense, the particular changes in the band structure introduced by the Li ions would be averaged out in the low-concentration limit. However, local quantities related to the Li ions, like the local density of states or electron forces in the vicinity of the impurity, would converge for higher $n$ values. Importantly, the same qualitative behavior is found for all considered positions in Fig.~\ref{fig:Li_positions}(a), as shown in Ref.~\cite{aguirre2026_rep}. In these cases, we observe good accuracy of the TB model for all considered supercell sizes.

Although not considered in this work, for the $n=1$ case (i.e., \ce{LiC6}), TB models are typically based on Kekulé-O-type modulated hopping amplitudes \cite{andrade2025,farjam2009} that depend on the distance between the \ce{C-C} bonds and the Li ion, which is consistent with the 380~meV energy gap observed in Ref.~\cite{bao2021}. Such hopping modulations, however, are driven by the high Li concentration of the composite system. In contrast, our model targets the opposite limit of low concentrations, where a simpler approximation based primarily on modifications to the carbon onsite energies is sufficient to capture the essential effects of the Li ion on the electronic band structure.

\begin{table}[h!]
  \centering
  \setlength{\tabcolsep}{9pt}
  \renewcommand{\arraystretch}{1.2}  
  \begin{tabular}{c|S[table-format=-2.3]|S[table-format=-2.3]|S[table-format=-2.3]}
    \hline \hline
    \multirow{2}{*}{\textbf{Fitting parameter}} & \multicolumn{3}{c}{\textbf{Supercell size $n$}} \\
    \cline{2-4}
    & {2} & {3} & {4} \\
    \hline \hline
    $\epsilon_\tf{Li}$  & -0.698 &  0.028 &  0.524 \\
    $\epsilon_0$        & -3.378 & -3.640 & -3.803 \\
    $\gamma_0$          & -3.373 & -2.193 &  0.451 \\
    $\sigma$            &  1.726 &  1.670 &  1.655 \\
    $\sigma_z$          &  2.112 &  1.967 &  1.942 \\
    $\tau$              &  2.320 &  2.197 & -6.505 \\
    $\tau_z$            &  2.926 &  5.418 & 14.484 \\
    \hline \hline
  \end{tabular}
  \caption{Obtained TB parameters for Li adsorbed on MLG. We use eV for $\epsilon_\tf{Li}$, $\epsilon_0$, and $\gamma_0$; and {\AA} for $\sigma$, $\sigma_z$, $\tau$, $\tau_z$, respectively.}
  \label{tab:monolayer}
\end{table}

In Fig.~\ref{fig:bands_mono_size}~(c) and (d), we show the dependence of the DFT and TB energy bands on the vertical distance between the Li ion and the MLG substrate. Similar to the discussion above, the modifications to the band structure diminish as the height of the Li ion increases, since the perturbation to the carbon onsite energies and the \ce{Li-C} hopping terms both weaken.

It is interesting to note that for a Li ion closer to the MLG layer, there is a pronounced electron-hole symmetry breaking, which diminishes for greater heights. Aligned with this trend, we observe an improvement in the accuracy of the TB bands. This is also observed for the remaining Li ion positions and supercell sizes, where the TB bands show better agreement when the electron-hole symmetry breaking is reduced, as shown in Ref.~\cite{aguirre2026_rep}.

The adjusted $\bm{\alpha}$-parameters defined in Eq.~\eqref{eq:parameters} for supercell sizes $n = 2$, 3, and 4 are listed in Table~\ref{tab:monolayer}. For each value of $n$, we considered the four in-plane positions from Fig.~\ref{fig:Li_positions}(a) and three vertical positions of the Li ion on MLG. As shown in the table, the relevant parameters ($\epsilon _0$, $\sigma$, and $\sigma_z$), which describe the onsite energy corrections in Eq.~\eqref{eq:Li-pot}, exhibit a small variation across the considered supercell sizes. To put these values into physical perspective, the spatial extent of the Li-ion potential, given by $\sigma$ and $\sigma_z$, is comparable to the nearest-neighbor \ce{C-C} distance. Furthermore, for position $P_1$ with $n=2$, the onsite energy modification of the C atoms closest to the Li ion is $\Delta\epsilon \sim -0.35 \, |t_0|$. The remaining parameters, related to the \ce{Li-C} hopping terms ($\gamma_0$, $\tau$, and $\tau_z$) and the Li onsite energy $\epsilon_\tf{Li}$, show a larger variation since their impact on the low-energy bands is considerably smaller. In other words, substantial fluctuations in these parameters have a negligible effect on the TB bands near the Fermi level. For this reason, in the case of monolayer graphene, the inclusion of a Li ion per supercell given by Eq.~\eqref{eq:Hint} can be simplified to:
\begin{equation}
\hat{H}_\tf{int} =  \sum_{ij} \sum_\alpha \epsilon(\bm{r}_{\alpha,i}-\bm{r}_{\tf{Li},j})  \hat{c}^\dagger_{\alpha,i} \hat{c}_{\alpha,i},
 \label{eq:simplified_Hint}
\end{equation}
i.e., we can neglect the \ce{Li-C} hopping terms. This interaction term is sufficient to reproduce the band behavior around the $\Gamma$ point shown in Fig.~\ref{fig:bands_mono_pos} with almost the same accuracy as the complete model. Although a minor difference between the predictions of the two models can be found for $P_1$ position, the \ce{Li-C} hopping terms can be disregarded for most applications, given the significant simplification.
The parameters for the simplified model are listed in Table~\ref{tab:simplified_monolayer}.

\begin{table}[h!]
  \centering 
  \setlength{\tabcolsep}{9pt}
  \renewcommand{\arraystretch}{1.2}
  \begin{tabular}{c|S[table-format=-2.3]|S[table-format=-2.3]|S[table-format=-2.3]}
    \hline \hline
    \multirow{2}{*}{\textbf{Fitting parameter}} & \multicolumn{3}{c}{\textbf{Supercell size $n$}} \\
    \cline{2-4} 
    & {2} & {3} & {4} \\
    \hline \hline
    $\epsilon_0$ & -3.561 & -3.664 & -3.887 \\
    $\sigma$     &  1.742 &  1.687 &  1.718 \\
    $\sigma_z$   &  2.051 &  1.981 &  1.981 \\
    \hline \hline
  \end{tabular}
  \caption{Obtained TB parameters for the simplified model of Li adsorbed on MLG. We use eV for $\epsilon_0$ and {\AA} for $\sigma$ and $\sigma_z$.}
  \label{tab:simplified_monolayer}
\end{table}

\begin{figure}[t!]
  \centering
  \includegraphics[width=1.0\linewidth]{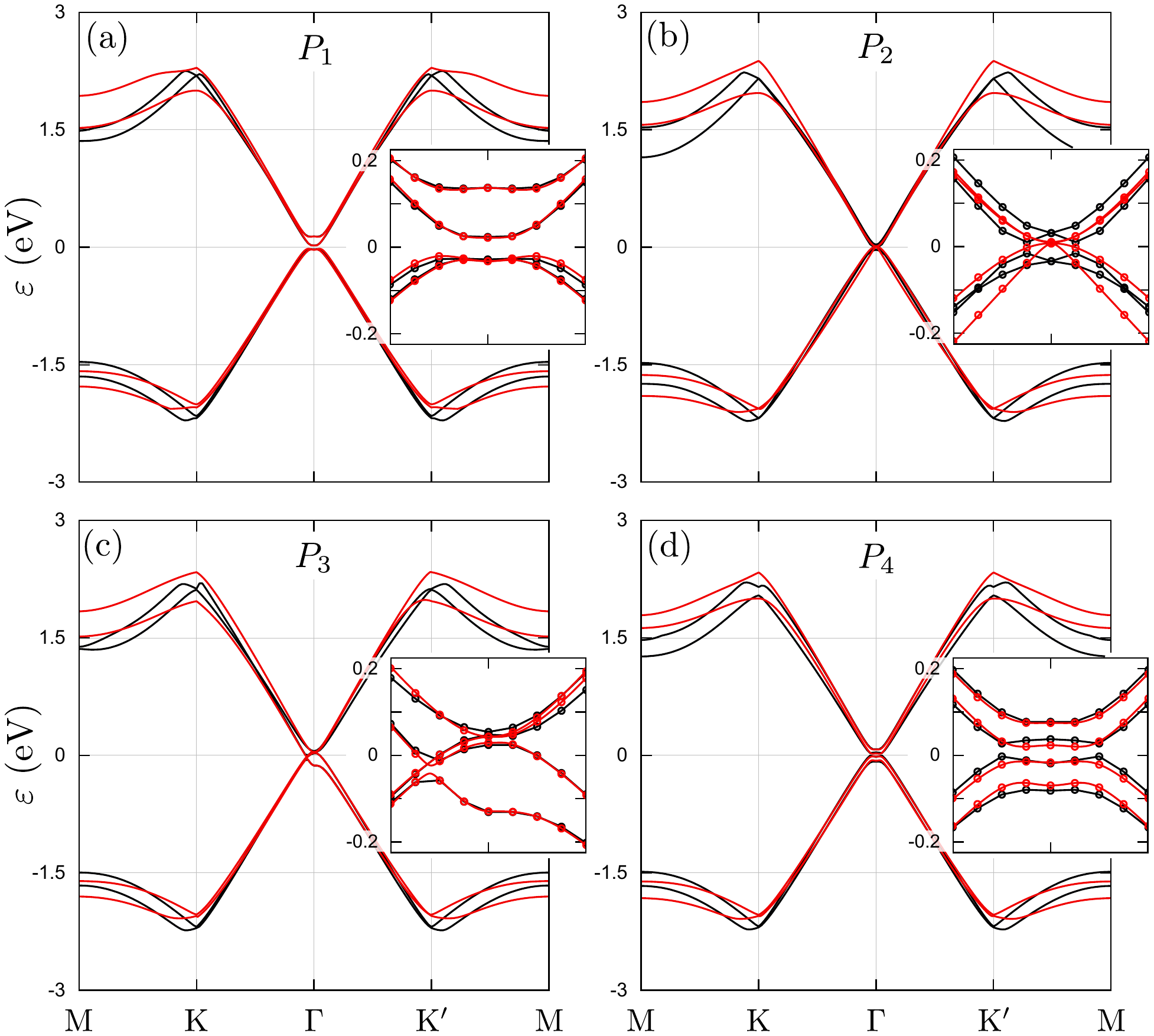}
  \caption{Band structure for Li-ion intercalated within BLG, relative to the Fermi level (Dirac point) of the pristine substrate. Here we take $n=2$ and the Li-ion positions depicted in Fig.~\ref{fig:Li_positions}(b).}
  \label{fig:bands_bi_pos}
\end{figure}

\textit{AB-stacked bilayer graphene}.
Fig.~\ref{fig:bands_bi_pos} shows the DFT-calculated electronic bands (black lines) for a Li ion in a supercell of BLG of $n = 2$ near the Fermi level, corresponding to the Li positions depicted in Fig.~\ref{fig:Li_positions}(b).
The bands obtained using the TB model described in Sec.~\ref{sec:TBmodel} for these positions are shown in red. The TB parameters were obtained by fitting the model to the DFT bands and are listed in Table~\ref{tab:bilayer} for Li in BLG with $n = 2$. As with MLG, the parameter set was selected by emphasizing accuracy near the Dirac point through a statistically weighted fitting procedure.

Here, we can again analyze the obtained band structure in terms of direct and indirect corrections of pristine BLG due to the local potential and hopping terms introduced by the Li ion. Importantly, although in pristine BLG we would have eight bands near the Dirac points, the interlayer coupling $t_1$ splits apart four of them, such that the effective $\mathcal{D}$ space is again fourfold degenerate. In this case, only the A1 and B2 sublattices contribute to the $\bm{h}_\mathcal{DD}$ matrix, while sublattices B1 and A2 are included in $\bm{\Sigma}_\mathcal{DD}$, see Appendix~\ref{app:effective_hamiltonian}.

Figure~\ref{fig:bands_bi_pos}(a) shows four electronic bands near the Dirac point for the lowest-energy configuration labeled $P_1$ in Fig.~\ref{fig:Li_positions}(b). Since this configuration exhibits both $x$ and $y$ mirror symmetries, the bands display a symmetric profile around the $\Gamma$ point along the chosen path. The presence of the Li ion opens a band gap and also shifts the two upper bands differently with respect to the two lower ones. Since A1 and B2 are the sublattices contributing to $\bm{h}_\mathcal{DD}$, we can see that $P_1$ shares the same symmetry as $P_2$ of MLG. In fact, the effective matrix $\bm{h}^\tf{eff}_\mathcal{DD}$ displays the same structure for both $P_1$-BLG and $P_2$-MLG, yielding two degenerate eigenenergies and another two nondegenerate ones. The relative order between the degenerate and nondegenerate energy levels depends on the intravalley $v$ and intervalley $w = w_\tf{x}$ couplings. While in $P_2$-MLG the degenerate energies fall between the nondegenerate ones---ensuring the closure of the gap in Fig.~\ref{fig:bands_mono_pos}(b)---in $P_1$-BLG we obtain $|v|>|w|$, and the degenerate energies fall below the nondegenerate ones, such that a gap opens at the supercell $\Gamma$ point. The figure also shows that the TB model reproduces the main features of the DFT bands. Near the $\Gamma$ point, the agreement is particularly good, as shown in the inset.

Position $P_2$ is shown in panel b, where the Li ion is located between two aligned C atoms from different graphene layers. For this position, the DFT bands exhibit a small gap that is not reproduced by the TB model using the parameter set from Table~\ref{tab:bilayer}. For energies close to the Fermi level, this position resembles that of $P_1$-MLG, since the Li ion is located exactly in the middle of the two layers, and the two sublattices associated with the C atoms aligned with the Li ion do not contribute to $\bm{h}_\mathcal{DD}$. Consequently, the direct correction only contains the average potential in its diagonal elements. Additionally, no significant corrections are observed from $\bm{\Sigma }_\mathcal{DD}$, so the four energies at the supercell $\Gamma$ point remain almost degenerate in the TB model.\footnote{Although the four eigenenergies split into two levels, their difference is negligible on the scale of the inset.} We hypothesize that this disagreement arises from a particular Li charge distribution profile that strongly differs from the other considered configurations; this effect would alter the local chemical environment but are not easily captured in the present non-self-consistent TB framework. Nonetheless, since this position corresponds to the highest-energy configuration in Fig.~\ref{fig:Li_positions}(b), this discrepancy is less critical due to its low thermodynamic probability.

Figure~\ref{fig:bands_bi_pos}(c) shows the bands for position $P_3$, where the Li ion is at the center of one of the rhombuses defined by the overlapping hexagons of the two graphene layers. As seen in the figure, the TB model reproduces most of the DFT band features, including the strong deformation around the $\Gamma$ point. Excluding the sublattices associated with the vertically aligned C atoms, this position resembles that of $P_3$-MLG, where the ``bond'' is defined by the C atoms aligned at 30$^\circ$ across the two layers. In this sense, the structure of $\bm{h}_\mathcal{DD}$ is similar for both positions, showing no direct intravalley coupling ($v=0$) and different intervalley couplings ($w \neq w_\tf{x}$). For BLG, however, no Dirac points are observed, since the lowest conduction and highest valence bands overlap around the $\Gamma$ point. Additionally, the broken $x$ and $y$ mirror symmetries of this configuration are mirrored by the asymmetric dispersion of the bands around the $\Gamma$ point.

In the case of position $P_4$ (panel d), the Li ion is situated below the midpoint of a \ce{C-C} bond of the top layer, with its vertical position located at $z =1.5858$~{\AA} (slightly closer to the bottom layer). The TB model, in general, remains in good agreement with the DFT bands. Around the supercell $\Gamma$ point, the Li ion opens a band gap, completely lifting the fourfold degeneracy. In this configuration, all coupling terms in $\bm{h}_\mathcal{DD}$ are non-zero and distinct; consequently, the direct correction alone is sufficient to account for the energy level splitting around the Fermi level.

\begin{table}[h!]
  \centering
  \setlength{\tabcolsep}{18pt}
  \renewcommand{\arraystretch}{1.2}
  \begin{tabular}{c|S[table-format=-2.3]|S[table-format=-2.3]}
    \hline \hline
    \multirow{2}{*}{\textbf{Fitting parameter}} & \multicolumn{2}{c}{\textbf{Supercell size $n$}} \\
    \cline{2-3}
    & {2} & {3} \\
    \hline \hline
    $\epsilon_\tf{Li}$  &  -3.044 &  -3.599 \\
    $\epsilon_0$        &  -3.370 &  -3.530 \\
    $\gamma_0$          & -18.541 & -23.853 \\
    $\sigma$            &   2.489 &   2.710 \\
    $\sigma_z$          &   1.541 &   1.684 \\    
    $\tau$              &   1.600 &   1.555 \\
    $\tau_z$            &   1.342 &   1.285 \\
    \hline \hline
  \end{tabular}
  \caption{Obtained TB parameters for Li intercalated within BLG. We use eV for $\epsilon_\tf{Li}$, $\epsilon_0$, and $\gamma_0$; and {\AA} for $\sigma$, $\sigma_z$, $\tau$, $\tau_z$, respectively.}
  \label{tab:bilayer}
\end{table}

The adjusted $\bm{\alpha}$-parameters of Eq.~\eqref{eq:parameters} for BLG supercells sizes $n = 2$ and 3 are tabulated in Table~\ref{tab:bilayer}.
As in MLG, the parameters ($\epsilon _0$, $\sigma$, and $\sigma_z$), which describe the onsite energy corrections in Eq.~\eqref{eq:Li-pot}, exhibit small variations between the two considered supercell sizes.
Furthermore, the $\epsilon _0$ parameter shows similar values to  those obtained for MLG and also decreases as the supercell size increases.
However, a comparison of the $\sigma$ and $\sigma_z$ parameters in MLG and BLG  shows that in MLG the $\sigma$ parameter is smaller than the $\sigma_z$ parameter, whereas in BLG the opposite relation is observed.
This means that in MLG, the spatial distribution of the correction to the carbon onsite energies corresponds to a prolate ellipsoidal shape, whereas in BLG it becomes oblate. This structural transition mimics an $s$-type orbital---initially spherical in isolation---deformed by its environment. Physically, the intercalation in BLG imposes vertical electrostatic confinement and symmetric screening between the carbon sheets, compressing the potential vertically. In contrast, the open surface of MLG lacks this constraint, allowing the localized dipole perpendicular to the layer to elongate the distribution along the $z$-axis.

In contrast to MLG, the parameters related to the \ce{Li-C} hopping terms ($\gamma_0$, $\tau$, and $\tau_z$) and the Li onsite energy $\epsilon_\tf{Li}$ show  small variations for different supercell sizes.
In particular,  the values of the $\gamma_0$ parameter alone suggest relatively large \ce{Li-C} hopping terms.
However, when these hoppings are calculated  using Eq.~\eqref{eq:explicit}, together with the $\tau$ and $\tau_z$ parameters of Table~\ref{tab:bilayer}, they exhibit values comparable to $t_0$ for the nearest C  neighbors in the relevant positions  shown in Fig.~\ref{fig:Li_positions}(b).
In the case of BLG, we were unable to find a simplified model in which the \ce{Li-C} hopping terms can be neglected while still reproducing the band structures of Fig.~\ref{fig:bands_bi_pos} with appropriate accuracy.

In the $n=3$ case (see Ref.~\cite{aguirre2026_rep}), the electronic bands exhibit the same qualitative features previously discussed for $n=2$ and shown in Fig.~\ref{fig:bands_bi_pos}.
We also observe  good agreement between the TB model and the reference calculations for this supercell size.
As observed for MLG, the energy scale at which these features are visible is reduced as $n$ increases  because the number of C atoms in the supercell that remain unaffected by the Li ion increases.

\section{Conclusions}
\label{sec:conclusions}

We have developed and validated a tight-binding model applicable to Li impurities either adsorbed on monolayer graphene or intercalated within AB-stacked bilayer graphene. Alongside its structural simplicity, a key strength of the proposed model is its ability to account for any general impurity position within the supercell. This capability was demonstrated across different supercell sizes and can be readily extended to a dilute concentration regime. The model parameters were obtained by fitting to DFT data for different Li positions, vertical distances, and supercell sizes. Overall, the model reproduces the main features of the electronic bands near the Fermi level with high accuracy, particularly in the vicinity of the Dirac points, which constitute the primary focus of the parametrization.

The analysis of the considered TB Hamiltonian reveals that the modifications induced by the Li ion are strongly dictated by its local environment and symmetry. Depending on the adsorption or intercalation position, the impurity can induce interconnected effects such as band-gap openings, valley mixing, sublattice-symmetry breaking, or Dirac cone shifts. These effects are consistently captured by the fitted TB model and can be interpreted in terms of direct and indirect corrections to the low-energy electronic states of the pristine substrate.

For monolayer graphene, we find that the dominant contribution arises from local modifications of the carbon onsite energies. The corresponding fitting parameters exhibit only a weak dependence on the supercell size, indicating that the impurity potential is highly localized. As a consequence, a simplified description that neglects the \ce{Li-C} hopping terms is sufficient to reproduce the low-energy electronic structure with nearly the same accuracy as the complete model. In contrast, for bilayer graphene the \ce{Li-C} hopping terms remain essential and cannot be neglected without a significant loss of precision. The fitted parameters also reveal systematic differences between monolayer and bilayer graphene in the spatial distribution of the impurity-induced potential, which is consistent with the different local environments experienced by the Li ion in the two systems.

Finally, in line with Refs.~\cite{wang2007,pollak2010,astles2024}, the calculated interaction energies indicate that \ce{Li-Li} interactions decay with increasing supercell size and are more strongly screened in bilayer graphene. The reduction of the impurity-induced energy scales with increasing dilution further supports the interpretation that the low-energy electronic modifications originate from localized perturbations around the Li ion. These results provide an efficient framework for describing dilute Li impurities in graphene-based systems and pave the way for future studies of local electronic properties, impurity-induced electron transport, and other impurity-driven phenomena in graphene and related materials.

\section*{Acknowledgments}

H.A. acknowledges PhD fellowship from CONICET.
H.L.C. acknowledges financial support from CONICET (Grant No. PIP 2022-59241), and SECyT-UNC through Grant No. 33620230100363CB. 
E.M.P. acknowledges financial support from SECyT-UNC through Grant No. 33820230100101CB and 33820250100145CB.
All authors are members of CONICET.

\section*{Data Availability}
The data that support the findings of this article are openly available at Ref.~\cite{aguirre2026_rep}. These datasets include the first-principles and tight-binding electronic band structures, the reciprocal-space $\bm{k}$-point grids, and the total energies for all considered monolayer and bilayer graphene supercell configurations.

\appendix

\section{Fitting Considerations}
\label{app:fitting}

\begin{figure}[t!]
\centering
\includegraphics[width=1.0\linewidth]{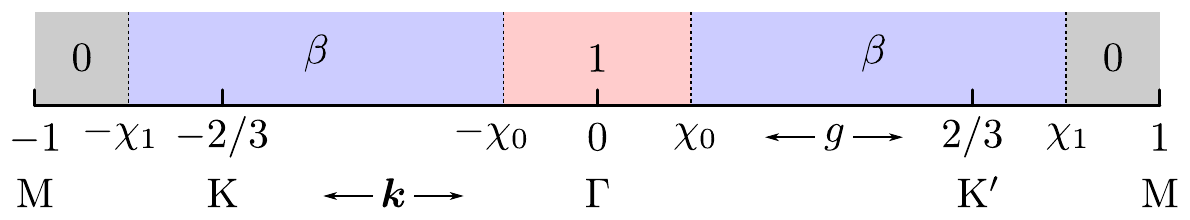}
\caption{
  Schematic representation of the weight function $\omega_n$ defined in Eq.~\eqref{eq:wn_f}, together with its explicit parametrization along the \ce{M-K-$\Gamma$-K$'$-M} path.
  Depending on the proximity to the $\Gamma$ point, the weight function takes the value of 1 (red), $\beta$ (blue), or 0 (gray).
}
\label{fig:weight}
\end{figure}

Fig.~\ref{fig:weight} shows a schematic representation of the weight function $\omega_n$ from Eq.~\eqref{eq:wn_f}, where the chosen $\bm{k}$-path is linearly parametrized in consecutive steps. Specifically, between the high-symmetry points $\bm{G}_i$ and $\bm{G}_j$, the path is given by:
\begin{equation}
\bm{k}(g) = \bm{G}_i + \frac{\bm{G}_j - \bm{G}_i}{g_j - g_i} (g - g_i), \quad g_i \leqslant g \leqslant g_j \,.
\end{equation}
In this scheme, the path of Fig.~\ref{fig:supercells}(b) is represented by a straight line that begins and ends at the $\tf{M}$ point. The corresponding $g$-values are also indicated in the figure.
Two regions around the $\Gamma$ point are defined by the parameters $\chi_0$ and $\chi_1$, respectively.
The first region, colored in red, applies a unit weight to the residuals. The second region, colored in blue, scales the residuals by the factor $\beta$.
This functional form of $\omega_n$ allows us to prioritize the reproduction of the DFT bands around the $\Gamma$ point within the TB model, while preserving the overall dispersion of these bands away from it.

The problem of multidimensional non-linear least-squares fitting (NLSF) requires the minimization of a cost function $\Phi_n$, as defined in Eq.~\eqref{eq:cost_function}.
This function collects the squared residuals between the TB model and the DFT bands, which depend directly on the TB parameters, $\bm{\alpha}$, as defined in Eq.~\eqref{eq:parameters}.
Within this framework, choosing appropriate initial values for $\bm{\alpha}$ represents a well-known challenge in NLSF, as these initial guesses strongly dictate the optimization outcome.
Furthermore, the parameters collected in $\bm{\alpha}$—which describe the energy and spatial distribution of the pseudoorbitals—are typically non-unique.
In this work, the criterion adopted to determine $\bm{\alpha}$ aims to obtain physically sound parameters with reasonable magnitudes for both energy and spatial distributions.

For each supercell size of MLG, the bands of twelve Li positions were 
simultaneously used to obtain the $\alpha$ parameters of Table~\ref{tab:monolayer}.
These positions correspond to the Li configurations of Fig.~\ref{fig:Li_positions}(a) on three different planes parallel to the MLG sheet.
One of these planes is at a distance of 1.6857~{\AA} from the MLG sheet and contains the lowest-energy configuration for the Li ion, which is found at $P_1$.
The other two planes are above and below this plane, at distances of 1.8857~{\AA} and 1.4857~{\AA} from the MLG sheet, respectively.

For BLG, the bands of six Li positions were simultaneously used to obtain the $\alpha$ parameters of Table~\ref{tab:bilayer} for each supercell size.
Four of these positions are shown in Fig.~\ref{fig:Li_positions}(b), where the lowest-energy configuration is also labeled $P_1$.
The other two positions, together with the initial guesses for $\bm{\alpha}$, the parameters for Eq.~\eqref{eq:wn_f}, and all the data used in each fitting procedure for both MLG and BLG, can be found in the repository available at Ref.~\cite{aguirre2026_rep}.

\section{Effective Hamiltonian at the Dirac point}
\label{app:effective_hamiltonian}

In this section we introduce the partition scheme for the fourfold degenerate subspace at the Dirac point. We begin by computing the Bloch Hamiltonian $\bm{H}_{\bm{k}}$ in either MLG or BLG pristine lattices, spanned by the primitive vectors:
\begin{equation}
\bm{a}_1 = a \, \bm{e}_{\pi/3}, \; \bm{a}_2 = a \, \bm{e}_{2\pi/3},
\end{equation}
with $a = \sqrt{3} a_0$ the lattice parameter and $\bm{e}_\theta = \cos\theta\bm{e}_x+\sin\theta\bm{e}_y$ the unit vector along the $\theta$-direction. The corresponding primitive vectors of the reciprocal lattice are thus given by:
\begin{equation}
\bm{b}_1 = \frac{4\pi}{\sqrt{3}a} \, \bm{e}_{\pi/6}, \; \bm{b}_2 = \frac{4\pi}{\sqrt{3}a} \, \bm{e}_{5\pi/3}.
\end{equation}
By taking the standard sublattice basis $\{\ket{\tf{A}},\ket{\tf{B}}\}$ for MLG and $\{\ket{\tf{A1}},\ket{\tf{B1}},\ket{\tf{A2}},\ket{\tf{B2}}\}$ for BLG, these Hamiltonian matrices write, respectively
\begin{align}
\bm{H}_{\bm{k}}^{\tf{MLG}} &= \begin{pmatrix}
0 & -t_0 f_{\bm{k}} \\
-t_0 f_{\bm{k}}^* & 0
\end{pmatrix},\\
\bm{H}_{\bm{k}}^{\tf{BLG}} &= \begin{pmatrix}
0 & -t_0 f_{\bm{k}} & 0 & 0 \\
-t_0 f_{\bm{k}}^* & 0 & t_1 & 0 \\
0 & t_1 & 0 & -t_0 f_{\bm{k}} \\
0 & 0 & -t_0 f_{\bm{k}}^* & 0
\end{pmatrix},
\end{align}
where $f_{\bm{k}}=1+\exp(i\bm{k}\cdot\bm{a}_1)+\exp(i\bm{k}\cdot\bm{a}_2)$. Notice that for BLG we only take in this analysis the interlayer coupling $t_1$ between vertically aligned carbon atoms. For a given wavevector $\bm{k}$, the MLG eigenenergies are the following:
\begin{equation}
\epsilon_s(\bm{k}) = (-1)^s t_0 |f_{\bm{k}}|,
\end{equation}
with $s = 1,2$ the band index. The corresponding eigenstates are the following:
\begin{equation}
\ket{\bm{k},s} = \frac{e^{i \phi_{\bm{k}}}\ket{\tf{A}}-(-1)^s \ket{\tf{B}}}{\sqrt{2}},
\end{equation}
with $\phi_{\bm{k}} = \tan^{-1}[\tf{Im}(f_{\bm{k}})/\tf{Re}(f_{\bm{k}})]$. Similarly, for BLG we have the following eigenenergies:
\begin{align}
\epsilon_{1,2}(\bm{k}) &= \mp \frac{t_1}{2}-\sqrt{\left(\frac{t_1}{2}\right)^2+t_0^2|f_{\bm{k}}|^2},\\
\epsilon_{3,4}(\bm{k}) &= \mp \frac{t_1}{2}+\sqrt{\left(\frac{t_1}{2}\right)^2+t_0^2|f_{\bm{k}}|^2},
\end{align}
and corresponding eigenstates,
\begin{align}
\ket{\bm{k},s} =& \frac{1}{\sqrt{2}\,\mathcal{N}_{\bm{k},s}} \! \left[ (-1)^s f_{\bm{k}}^* \ket{\tf{A1}}-(-1)^s\frac{\epsilon_s(\bm{k})}{t_0} \ket{\tf{B1}} \right. \nonumber  \\
 & \left. - \frac{\epsilon_s(\bm{k})}{t_0} \ket{\tf{A2}} + f_{\bm{k}} \ket{\tf{B2}}  \right],
\end{align}
with $s=1,2,3,4$ the band index, and
\begin{equation}
\mathcal{N}_{\bm{k},s} = \sqrt{\left(\frac{\epsilon_s(\bm{k})}{t_0}\right)^2 + |f_{\bm{k}}|^2} \, ,
\end{equation}
the corresponding normalization factors. To account for the Li ion on the substrate, we expand the above solutions across the supercell, yielding
\begin{equation}
\ket{\psi_{\bm{k},s}} = \frac{1}{\sqrt{N_\tf{c}}}\sum_{\bm{R},\ell} e^{i \bm{k} \cdot \bm{R}} \braket{\ell|\bm{k},s}\ket{\bm{R}+\bm{r}_\ell},
\end{equation}
where $N_\tf{c} = 3n^2$ is the number of pristine unit cells contained within the $(\sqrt{3}n\times\sqrt{3}n) \, \tf{R} \, 30^\circ$ supercell. The index $\ell$ runs over the sublattice basis, whose positions relative to the lattice vector $\bm{R}$ are given by $\bm{r}_\ell$, and the summation over $\bm{R}$ spans the supercell area.

The primitive vectors of the $n$-size supercell, on the other hand, are given by
\begin{equation}
\bm{A}_1 = 3na_0 \, \bm{e}_{\pi/6}, \; \bm{A}_2 = 3na_0 \, \bm{e}_{5\pi/6}, 
\end{equation}
such that the corresponding reciprocal lattice is spanned by the primitive vectors:
\begin{equation}
\bm{B}_1 = \frac{4\pi}{3\sqrt{3}na_0} \, \bm{e}_{\pi/3}, \; \bm{B}_2 = \frac{4\pi}{3\sqrt{3}na_0} \, \bm{e}_{2\pi/3},
\end{equation}
and we can take for the valleys: $\bm{K}=n\bm{B}_1$ and $\bm{K}'=n\bm{B}_2$. When evaluating the eigenenergies at the supercell $\Gamma$ point, the folding problem thus reduces to account for the $N_\tf{c}$ reciprocal vectors $\bm{G} = m_1 \bm{B}_1+m_2\bm{B}_2$, with $m_i \in\mathbb{Z}$, that fall within the FBZ of the pristine lattice. By evaluating $\bm{k}$ and $s$ at these points, we construct the total Hilbert space $\mathcal{H}$ basis. For MLG, the dimension of $\mathcal{H}$ is $2\times3n^2+1$, since we have two bands per $\bm{G}$-point and we include the Li-ion orbital. Consistently, for BLG we have $\dim \mathcal{H} = 4\times3n^2+1$. From these spaces, we denote as $\mathcal{H}_\mathcal{D}$ the fourfold degenerate subspace at the $\Gamma$ point and as $\mathcal{H}_\mathcal{C}$ its complementary subspace. We then proceed by writing the eigenvalue equation for the total Bloch Hamiltonian of the supercell, including the Li ion:
\begin{equation}
\begin{pmatrix}
\bm{h}_{\mathcal{DD}} & \bm{h}_\mathcal{DC} \\
\bm{h}_{\mathcal{CD}} & \bm{h}_\mathcal{CC}
\end{pmatrix}
\begin{pmatrix}
\bm{\psi}_{\mathcal{D}} \\
\bm{\psi}_{\mathcal{C}}
\end{pmatrix} = \epsilon
\begin{pmatrix}
\bm{\psi}_{\mathcal{D}} \\
\bm{\psi}_{\mathcal{C}}
\end{pmatrix} .
\end{equation}
Solving the second equation for $\bm{\psi}_\mathcal{C}$ and replacing it in the first equation yields the following effective Hamiltonian in $\mathcal{H}_\mathcal{D}$:
\begin{equation}
\bm{h}_{\mathcal{DD}}^\tf{eff}(\epsilon) = \bm{h}_{\mathcal{DD}}+\bm{h}_{\mathcal{DC}} (\epsilon-\bm{h}_{\mathcal{CC}})^{-1} \bm{h}_{\mathcal{CD}}.
\end{equation}
To compute these block matrices, we fix the value of $\epsilon$ at the average potential $\bar{\Delta}$ (see below) and calculate their matrix elements in the $\{\ket{\psi_{\bm{k},s}}\}$ basis. For MLG, we obtain
\begin{equation}
\bra{\psi_{\bm{k}',s'}}\!\hat{H}\!\ket{\psi_{\bm{k},s}} = e^{i\phi_{\bm{k},\bm{k}'}}V_\tf{A}(\bm{q})+(-1)^{s+s'}V_\tf{B}(\bm{q}),
\end{equation}
where $\phi_{\bm{k},\bm{k}'} = \phi_{\bm{k}}-\phi_{\bm{k}'}$, $\bm{q}=\bm{k}-\bm{k}'$, and
\begin{equation}
    V_\ell(\bm{q}) = \frac{1}{2N_\tf{c}}\sum_{\bm{R}} e^{i\bm{q}\cdot\bm{R}}\Delta\epsilon(\bm{R}+\bm{r}_\ell),
\end{equation}
with $\Delta\epsilon$ the onsite energy correction due to the Li ions, see Eq.~\eqref{eq:Li-pot}. Similarly, for the coupling between the carbon atoms and the Li ions, we have
\begin{equation}
\bra{\ce{Li}}\!\hat{H}\!\ket{\psi_{\bm{k},s}} = e^{i\phi_{\bm{k}}}\gamma_{\tf{A}}(\bm{k}) - (-1)^{s}\gamma_{\tf{B}}(\bm{k}),
\end{equation}
where
\begin{equation}
    \gamma_\ell(\bm{k}) = \frac{1}{\sqrt{2N_\tf{c}}}\sum_{\bm{R}}e^{i\bm{k}\cdot\bm{R}} \sum_j \gamma(\bm{R} + \bm{r}_\ell - \bm{r}_{\ce{Li},j}),
\end{equation}
with $\bm{r}_{\ce{Li},j}$ the position of the Li ion on the $j$-th supercell and $\gamma$ the \ce{Li-C} hopping term given in Eq.~\eqref{eq:explicit}. Indeed, since this term decays exponentially with the bond length, only nearest-neighbor supercells need to be considered.

Focusing on the $\mathcal{D}$ subspace, we obtain for the diagonal elements of $\bm{h}_{\mathcal{DD}}$:
\begin{equation}
\bra{\psi_{\bm{k},s}}\!\hat{H}\!\ket{\psi_{\bm{k},s}} = \bar{\Delta} = \frac{1}{2N_\tf{c}} \sum_{\bm{R},\ell} \Delta\epsilon(\bm{R}+\bm{r}_\ell),
\end{equation}
which accounts for the supercell's average potential introduced by the Li ions. Notice that this result is independent of the chosen valley. For the band coupling on a given valley we have:
\begin{equation}   \bra{\psi_{\bm{k},\bar{s}}}\!\hat{H}\!\ket{\psi_{\bm{k},s}} = \Delta V = V_{\tf{A}}(\bm{0})-V_{\tf{B}}(\bm{0}),
\end{equation}
where $\bar{s}$ denotes the band index opposite to $s$ , i.e., $\bar{s}=2,1$ for $s=1,2$. Interestingly, this term is sensitive to the breaking of the sublattice symmetry due to the position of the Li ion.

\begin{figure}[t!]
  \centering
  \includegraphics[width=0.6\linewidth]{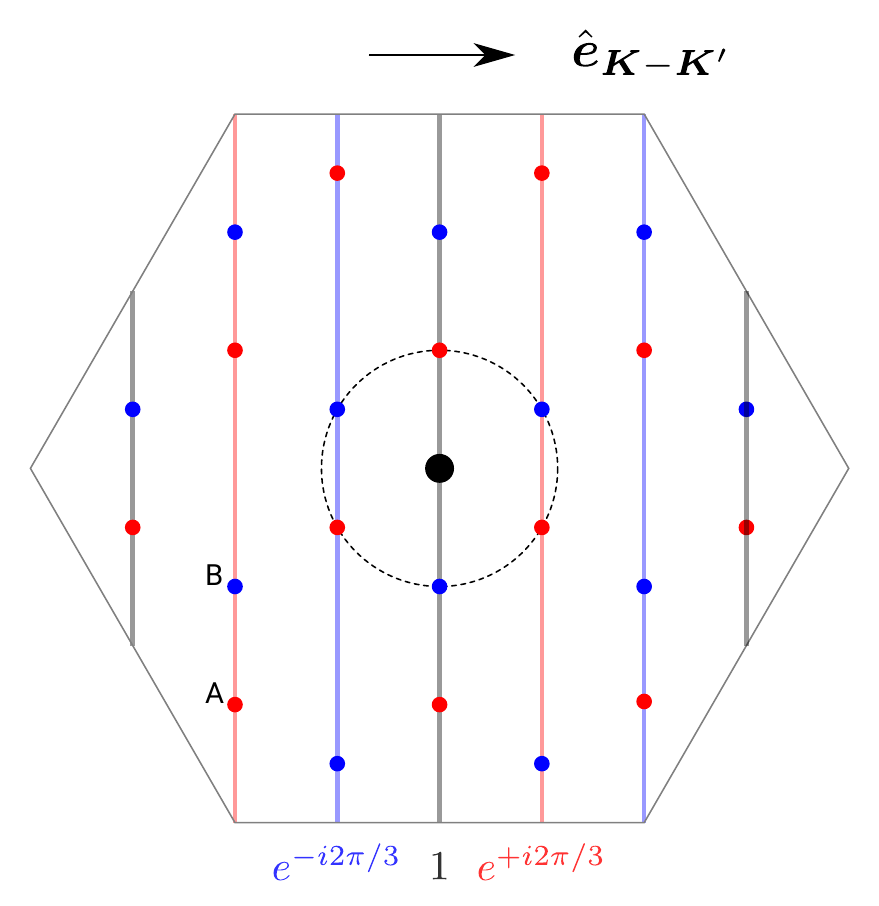}
  \caption{Direct intervalley coupling in MLG for the Li ion (black circle) at position $P_1$. Vertical lines show the phase factors $e^{i (\bm{K}-\bm{K}')\cdot \bm{R}}$, while red and blue circles represent C atoms belonging to the A and B sublattices, respectively.}
  \label{fig:MLG_phases}    
\end{figure}

In addition, we can consider the direct intervalley coupling in $\bm{h}_\mathcal{DD}$. 
Since the phase $\phi_{\bm{k}}$ is ill-defined exactly at the valley centers, we evaluate 
it at $\bm{k} = \bm{G} + \delta \, \bm{e}_\theta$, where $\bm{G} = \bm{K}$ or $\bm{K}'$. 
In the limit $\delta \to 0^+$, this parametrization yields $\phi_{\bm{K}} = -\theta$ and 
$\phi_{\bm{K}'} = \theta + \pi$. By choosing the gauge $\theta = -\pi/2$, the matrix 
elements reduce to:
\begin{equation}
\bra{\psi_{\bm{K}',s'}}\!\hat{H}\!\ket{\psi_{\bm{K},s}} = V_{\tf{A}}(\bm{q}) + (-1)^{s+s'} V_{\tf{B}}(\bm{q}),
\end{equation}
where $\bm{q} = \bm{K} - \bm{K}'$. For our choice where $\bm{q}\cdot\bm{R} = \bm{q}\cdot (n_1\bm{a}_1+n_2\bm{a}_2) = 2\pi/3(n_1-n_2)$, we obtain that sublattice contributions $V_\ell(\bm{q})$ may cancel out if there are a multiple of three C atoms equidistant to the Li ion such they have the three different phases, i.e., $n_1-n_2 = 0,1,2$. For the Li ion at position $P_1$ (hollow), shown in Fig.~\ref{fig:MLG_phases}, for each lattice vector $\bm{R}$ there is a pair of other lattice vectors $\bm{R}'$ such that $\Delta\epsilon(\bm{R}+\bm{r}_\ell) = \Delta\epsilon(\bm{R}'+\bm{r}_\ell)$, and the contributions from these points cancel out, yielding no direct valley mixing in this case. This is shown in the figure, where the contributions from the surrounding C atoms (central dashed line circle) interfere destructively within the A and B sublattices separately. In the position $P_2$ (top), the potential acts asymmetrically on the two sublattices. While the contributions from one sublattice might cancel out due to the relative phases, the sublattice directly coordinated with the Li ion provides a non-vanishing contribution, thus allowing for direct valley mixing.

We can proceed in a similar way for BLG, where the fourfold degenerate subspace $\mathcal{D}$ is spanned by \{$\ket{\bm{K},2}$, $\ket{\bm{K},3}$, $\ket{\bm{K}',2}$, $\ket{\bm{K}',3}$\} to compute $\bm{h}_\mathcal{DD}$. Setting the interlayer couplings $t_3=t_4=0$, the matrix elements read:
\begin{equation}
\bra{\psi_{\bm{k}',s'}}\!\hat{H}\!\ket{\psi_{\bm{k},s}} = (-1)^{s+s'}V_\tf{A1}(\bm{q})+V_\tf{B2}(\bm{q}),
\end{equation}
where $\bm{q}=\bm{0}$ for intravalley coupling and $\bm{q}=\bm{K}-\bm{K}'$ for intervalley coupling. In addition, we use the fact that the states in $\mathcal{D}$ have no weight on the B1 and A2 sublattices, since they are pushed away from the Dirac point due to the interlayer coupling $t_1$.

\bibliography{cite}

\end{document}